\documentclass[usenatbib]{mn2e}

\usepackage{graphicx}
\usepackage[usenames]{color}

\newlength{\colwidth}
\newcommand{\Msolyrkpcsq}{\mbox{M}_{\sun}\,\mbox{yr}^{-1}\,\mbox{kpc}^{-2}}
\newcommand{\Msolpcsq}{\mbox{M}_{\sun}\,\mbox{pc}^{-2}}
\newcommand{\Msolh}{h^{-1}\,\mbox{M}_{\odot}}
\newcommand{\Myr}{\mbox{Myr}}
\newcommand{\kpch}{h^{-1}\,\mbox{kpc}}
\newcommand{\pch}{h^{-1}\,\mbox{pc}}

\newcommand{\K}{\mbox{K}}

\newcommand{\cm}{{\rm cm}}

\newcommand{\kms}{{\rm km}\,{\rm s}^{-1}}

\newcommand{\erg}{{\rm erg}}
\newcommand{\yr}{{\rm yr}}

\newcommand{\Msun}{{{\rm M}_\odot}}

\begin{document}

\title[Galactic outflows with kinetic SN feedback]{Simulating galactic
  outflows with kinetic supernova feedback}

\author[C. Dalla Vecchia \& J. Schaye]
{Claudio Dalla Vecchia$^1$\thanks{E-mail: caius@strw.leidenuniv.nl} and
Joop Schaye$^1$\thanks{E-mail: schaye@strw.leidenuniv.nl}\\
$^1$Leiden Observatory, Leiden University, P.O. Box 9513, 2300 RA Leiden,
  the Netherlands
}

\maketitle

\abstract
Feedback from star formation is thought to play a key role in the
formation and evolution of galaxies, but its implementation in
cosmological simulations is currently hampered by a lack of numerical
resolution. We present and test a sub-grid recipe to model feedback
from massive stars in cosmological smoothed particle hydrodynamics
simulations. The energy is distributed in kinetic form among the gas
particles surrounding recently formed stars. The impact of the
feedback is studied using a suite of high-resolution simulations of
isolated disc galaxies embedded in dark halos with total mass
$10^{10}$ and $10^{12}~\Msolh$. We focus in particular on the effect
of pressure forces on wind particles within the disc, which we turn
off temporarily in some of our runs to mimic a recipe that has been
widely used in the literature. We find that this popular recipe gives
dramatically different results because (ram) pressure forces on
expanding superbubbles determine both the structure of the disc and
the development of large-scale outflows. Pressure forces exerted by
expanding superbubbles puff up the disc, giving the dwarf galaxy an
irregular morphology and creating a galactic fountain in the massive
galaxy. Hydrodynamic drag within the disc results in a strong increase
of the effective mass loading of the wind for the dwarf galaxy, but
quenches much of the outflow in the case of the high-mass galaxy.
\endabstract

\keywords
methods: numerical --- ISM: bubbles --- ISM: jets and outflows ---
galaxies: evolution --- galaxies: formation --- galaxies: ISM 
\endkeywords

\section{Introduction}

Core-collapse supernovae (SNe) and winds from massive stars feed back
energy into the interstellar medium (ISM). The energy released by
massive stars can efficiently suppress subsequent star formation by
destroying dense, star forming clouds, by generating supersonic
turbulence and, if the star formation density is
sufficiently high for individual supernova remnants to overlap, by
blowing gas out of the disc. In starburst galaxies
feedback from star formation may result in the development of a
galaxy-wide superwind which may (temporarily) remove a large fraction
of the gas, while in galaxies with less intense star formation
feedback may lead to the development of a galactic fountain. Because
these feedback mechanisms operate 
on time scales that are very short compared to the age of the
universe, they can lead to self-regulation. 

Feedback from star formation is thought to be a crucial ingredient for
models of the formation and evolution of galaxies. Without it, star
formation becomes much more efficient than observed, particularly in
low-mass galaxies \cite[e.g.][]{White&Frenk1991}. Among other things,
galactic winds are also thought 
to be responsible for the enrichment of the intergalactic medium with
heavy elements \cite[e.g.][]{Aguirre2001} and for pre-heating the gas
that ends up in groups and clusters of galaxies
\cite[e.g.][]{Ponman1999}.  

Modelling SN feedback in simulations of the formation of
galaxies is known to be a difficult task, mostly
because the injected thermal energy tends to be radiated away well
before it has any hydrodynamical effect
\citep[e.g.][]{Katz1996,Balogh2001}. This
overcooling problem is 
probably caused by the fact that current state-of-the-art simulations
still lack the resolution to capture the physics of the multiphase
ISM \cite[e.g.][]{Ceverino&Klypin2007}.

A typical SN ejects $\sim 1~\Msun$ at $\sim 10^4~\kms$, which
corresponds to a kinetic energy $\sim 10^{51}~\erg$. Since the
sound-crossing time is initially much smaller than the radiative
cooling time, the remnant starts out as an adiabatic blast
wave. Once radiative losses become important, the energy-conserving
blast wave gives way to a momentum conserving snow-plough phase, whose
deceleration is determined mostly by the amount of mass that is swept
up by the wind. The initial phases in the evolution of 
a superbubble
driven by multiple SNe are very similar to the evolution of an
individual SN remnant. However, if a superbubble
blows out of the disc, its subsequent evolution may be strongly
affected by the ram pressure of infalling gas and the
gravitational attraction of the galaxy 
\cite[e.g.][]{Silich1998,Maclow1999,Dercole1999,Silich2001,Fujita2004,Dubois2008}.

Current simulations lack the
resolution to resolve 
individual SNe or, in the case of cosmological simulations\footnote{By
  cosmological simulations we mean simulations that model the
  evolution of a representative volume of the universe, as opposed to
  simulations that zoom in on one or a few galaxies.} 
even superbubbles. The energy released by dying stars in a single
resolution element per time step is typically
distributed over a mass that exceeds the mass of SN ejecta
by many orders of magnitude. The initial expansion velocity and
post-shock temperature are therefore underestimated by large
factors. Since the radiative cooling time scales as $t_c \propto
T^{1/2}$ above 1~keV, this implies that the cooling time is greatly
underestimated. Because the injection radius is also far too large,
the initial cooling time tends to be smaller than the bubble
sound-crossing time. Thus, the simulation will skip the adiabatic
blast-wave phase and the energy will typically be radiated away before
a significant fraction of the thermal energy has been converted into
kinetic energy. 

The fact that poor resolution results in inefficient
thermal feedback is generally attributed to a lack of spatial
resolution: real SNe explode in hot bubbles of low-density gas, 
whereas the ISM in cosmological simulations consists of a single
dense, ``warm'' phase. Mass resolution is, however,
more fundamental. Without sufficient mass resolution, the first SNe
will not be able to create a hot, low-density ISM phase in the first
place. 

Cosmological simulations must resort to sub-grid recipes to solve the
overcooling problem.\footnote{In the absence of efficient feedback
  mechanisms, the amount of cold gas predicted by cosmological
  simulations is limited by resolution. Hence, it is possible to
  roughly match the observed amount of cold gas at a particular
  redshift by tuning the resolution.}
Two types of solutions appear to work: injecting
(part of) the SN energy in kinetic rather than thermal form
\cite[e.g.][]{Navarro1993,Mihos&Hernquist1994,Kawata2001,Kay2002,Springel2003,Oppenheimer2006,Dubois2008}
and/or suppressing radiative cooling by hand 
\cite[e.g.][]{Gerritsen1997,Mori1997,Thacker2000,Kay2002,Sommer-Larsen2003,Brook2004,Stinson2006}. Although
the suppression of cooling enables the efficient conversion of
thermal energy to kinetic energy, the maximum wind velocity will still
be underestimated if the total mass of the neighboring
resolution elements exceeds that of a superbubble. Kinetic feedback
can alleviate this problem by kicking only a small fraction of the
resolution elements near the star particle. Kinetic feedback thus
gives us the freedom to distribute a fixed amount of kinetic energy
over a varying amount of mass. However, using this freedom to increase
the initial wind velocity has the drawback that the
imposed winds become more poorly sampled and therefore less
isotropic. With increasing resolution, the two types of sub-grid models
for the generation of galactic winds are expected to converge.

A third approach, which is often combined with one of the above, is
to employ a sub-grid model to describe the multiphase ISM
\cite[e.g.][]{Yepes1997}. Examples 
include imposing an effective equation of state which specifies the
total pressure of the ISM as a function of its mean density
\cite[e.g.][]{Springel2003} and, at
least for the case of smoothed particle hydrodynamics
simulations, explictly decoupling thermal phases by
  using different 
types of particles to represent the hot and cold gas phases
\cite[e.g.][]{Marri&White2003,Scannapieco2006}. While galactic winds
can in principle be triggered naturally if the latter method is used,
this again requires ad-hoc solutions in the absence of sufficient
resolution.  

To prevent the overcooling problem, simulations that impose an
effective equation of state for 
the ISM must either make it extremely stiff, resulting in discs that
are much thicker and smoother than observed, or employ a sub-grid
recipe for galactic winds. However, even if it does not directly
generate winds, the use of 
an effective equation of state for dense gas can be
considered a necessary 
ingredient for simulations that lack the resolution and/or physics to
model the multiphase ISM. If the equation of state is not modified by hand,
gas will accumulate at unrealistically 
low temperatures and high densities in the absence of resolved
feedback processes. As 
we discussed in \cite{Schaye2008}, using a power-law equation of state
with a polytropic index of $4/3$ has the advantage 
of yielding a Jeans mass that is independent of the density which
makes it possible to suppress spurious fragmentation due to a lack of
resolution. 

It is, however, 
important to note that maintaining an effective equation of state
in the presence of radiative
losses would, in reality, require energy. The amount of energy
that is required depends on unresolved physical 
processes and can therefore not be reliably determined.
Imposing an equation of state for dense gas therefore implies that the
energy available for any wind sub-grid model must be less than the energy
provided by star formation. 
 
The most widely used recipe for galactic winds in SPH simulations is
the kinetic feedback model implemented by \cite{Springel2003} (hereafter
 SH03).  Motivated by
the desire to impose the net mass loading and velocity of the wind
after it has escaped from the disc and by the wish to make the recipe
insensitive to numerical 
resolution, hydrodynamical forces on the
wind particles are temporarily switched off in the SH03 model. Thus,
winds cannot blow bubbles in the disc, drive turbulence or 
create channels in gas with densities typical of the ISM. Their sole
effect on the disc is the removal of fuel for star formation by the
ejection of wind particles. 

Another aspect of the SH03 recipe is that
wind particles are selected stochastically from all the dense
(i.e.\ star-forming) particles in the simulation and are therefore
not constrained to be neighbours of newly-formed stars. While this
non-local feedback solves some numerical problems, we will argue that
it can lead to undesirable behavior in galaxies that do not contain
large numbers of particles (i.e.\ most galaxies in cosmological
simulations), particularly if metal enrichment is included. 

We present and test a modified a variation of the SH03 recipe in
which the winds are local and not decoupled hydrodynamically. In this
paper we will focus on the
effects of hydrodynamical drag within the ISM. Using high-resolution
SPH simulations of isolated disc galaxies we show that pressure forces
exerted by and on wind particles have a dramatic effect on both the
structure of the ISM and the 
development of galactic winds. Hydrodynamical drag on wind particles
converts low-mass disc galaxies into irregulars and results in a
strong increase of the net wind mass loading,
while for high-mass galaxies it leads to the creation
of bubbles and the development of a galactic
fountain. Pressure forces within the disc reduce the kinetic
energy in the wind above the disc by orders of magnitude.

This paper is organized as follows. We present our recipe for galactic
winds and its implementation in SPH simulations in
section~\ref{sec:recipe}. After describing our simulations of isolated
disc galaxies in
section~\ref{sec:sims}, we present the results on the galaxies'
morphology, star 
formation histories and large-scale winds in sections
\ref{sec:morphology} to \ref{sec:winds}, respectively. In section
\ref{sec:resol} we present resolution tests which show that
the effects of hydrodynamical drag are underestimated if the Jeans
scale is unresolved. Finally, we summarize and discuss our conclusions
in section~\ref{sec:disc}.

\begin{table*}
\begin{center}
\caption{Simulation parameters:
  total mass, $M$;
  input mass loading, $\eta$;
  input wind velocity, $v_{\rm w}$;
  total number of particles, $N_{\rm tot}$;
  total number of gas particles in the disc, $N_{\rm disc}$;
  mass of baryonic particles, $m_{\rm b}$;
  mass of dark matter particles, $m_{\rm DM}$;
  gravitational softening of baryonic particles, $\epsilon_{\rm b}$;
  gravitational softening of dark matter particles, $\epsilon_{\rm DM}$;
  wind feedback included, (Wind);
  wind particles hydrodynamically decoupled, (Decoupled). Values
  different from the fiducial ones are shown in bold.
\label{tbl:params}}
\begin{tabular}{cccccccccccc}

\hline
  Simulation & $M_{\rm halo}$ & $\eta$ & $v_{\rm w}$ & $N_{\rm tot}$ &
  $N_{\rm disc}$ & $m_{\rm b}$ & $m_{\rm DM}$ & $\epsilon_b$ &
  $\epsilon_{\rm DM}$ & Wind & Decoupled\\
             & $(\Msolh)$      &        & $(\kms)$     &              &              & $(\Msolh)$   & $(\Msolh)$    & $(\pch)$       & $(\pch)$             &     &       \\
\hline
  \textit{m10}        & $10^{10}$ & 2 & 600 & 5000$\,$494 & 235$\,$294 & $5.1\times 10^2$ & $2.4\times 10^3$ & 10 &  17 & Y & N \\
  \textit{m10nowind}  & $10^{10}$ & -- & -- & 5000$\,$494 & 235$\,$294 & $5.1\times 10^2$ & $2.4\times 10^3$ & 10 &  17 & \textbf{N} & -- \\
  \textit{m10$\eta$1v848} & $10^{10}$ & \textbf{1} & \textbf{848} & 5000$\,$494 & 235$\,$294 & $5.1\times 10^2$ & $2.4\times 10^3$ & 10 &  17 & Y & N \\
  \textit{m10$\eta$4v424} & $10^{10}$ & \textbf{4} & \textbf{424} & 5000$\,$494 & 235$\,$294 & $5.1\times 10^2$ & $2.4\times 10^3$ & 10 &  17 & Y & N \\
  \textit{m10dec}     & $10^{10}$ & 2 & 600 & 5000$\,$494 & 235$\,$294 & $5.1\times 10^2$ & $2.4\times 10^3$ & 10 &  17 & Y & \textbf{Y} \\
\hline
  \textit{m10lr008}   & $10^{10}$ & 2 & 600 &  \textbf{625$\,$061} &  \textbf{29$\,$411} & $\mathbf{4.1\times 10^3}$ & $\mathbf{1.9\times 10^4}$ & \textbf{20} &  \textbf{34} & Y & N \\
  \textit{m10lr064}   & $10^{10}$ & 2 & 600 &   \textbf{78$\,$132} &   \textbf{3$\,$676} & $\mathbf{3.3\times 10^4}$ & $\mathbf{1.5\times 10^5}$ & \textbf{40} &  \textbf{68} & Y & N \\
  \textit{m10lr512}   & $10^{10}$ & 2 & 600 &    \textbf{9$\,$766} &    \textbf{459} & $\mathbf{2.6\times 10^5}$ & $\mathbf{1.2\times 10^6}$ & \textbf{80} & \textbf{136} & Y & N \\
\hline
  \textit{m12}        & $10^{12}$ & 2 & 600 & 5000$\,$494 & 235$\,$294 & $5.1\times 10^4$ & $2.4\times 10^5$ & 10 &  17 & Y & N \\
  \textit{m12nowind}  & $10^{12}$ & -- & -- & 5000$\,$494 & 235$\,$294 & $5.1\times 10^4$ & $2.4\times 10^5$ & 10 &  17 & \textbf{N} & -- \\
  \textit{m12$\eta$1v848} & $10^{12}$ & \textbf{1} & \textbf{848} & 5000$\,$494 & 235$\,$294 & $5.1\times 10^4$ & $2.4\times 10^5$ & 10 &  17 & Y & N \\
  \textit{m12$\eta$4v424} & $10^{12}$ & \textbf{4} & \textbf{424} & 5000$\,$494 & 235$\,$294 & $5.1\times 10^4$ & $2.4\times 10^5$ & 10 &  17 & Y & N \\
  \textit{m12dec}     & $10^{12}$ & 2 & 600 & 5000$\,$494 & 235$\,$294 & $5.1\times 10^4$ & $2.4\times 10^5$ & 10 &  17 & Y & \textbf{Y}\\
\hline
  \textit{m12lr008}   & $10^{12}$ & 2 & 600 &  \textbf{625$\,$061} &  \textbf{29$\,$411} & $\mathbf{4.1\times 10^5}$ & $\mathbf{1.9\times 10^6}$ & \textbf{20} &  \textbf{34} & Y & N \\
  \textit{m12lr064}   & $10^{12}$ & 2 & 600 &   \textbf{78$\,$132} &   \textbf{3$\,$676} & $\mathbf{3.3\times 10^6}$ & $\mathbf{1.5\times 10^7}$ & \textbf{40} &  \textbf{68} & Y & N \\
  \textit{m12lr512}   & $10^{12}$ & 2 & 600 &    \textbf{9$\,$766} &    \textbf{459} & $\mathbf{2.6\times 10^7}$ & $\mathbf{1.2\times 10^8}$ & \textbf{80} & \textbf{136} & Y & N \\
\hline
\end{tabular}
\end{center}
\end{table*}

\section{Kinetic feedback implementation}
\label{sec:recipe}

Following \cite{Aguirre2001} and SH03, we specify the kinetic
feedback through two 
parameters: the initial mass loading and wind velocity. It is
convenient to introduce a dimensionless wind mass loading parameter
$\eta$ by expressing the initial wind mass loading $\dot{M}_{\rm w}$
in units of the  star formation rate $\dot{M}_\ast$, 
\begin{equation}
\dot{M}_{\rm w} \equiv \eta \dot{M}_\ast.
\end{equation}
Assuming that core-collapse SNe inject a kinetic energy of
$\epsilon_{\rm SN}$ per solar mass of stars formed, the fraction of this
energy carried by the wind is
\begin{eqnarray}
f_{\rm w} &=& {\eta v_{\rm w}^2 \over 2 \epsilon_{\rm SN}}  \\
& \approx & 0.4 \left ({\eta \over 2}\right ) \left ({v_{\rm w} \over
  600~\kms}\right )^2 \left ({\epsilon_{\rm SN} \over 1.8\times
  10^{49}~\erg\,\Msun^{-1}} \right )^{-1},
\end{eqnarray}
where $v_{\rm w}$ is the input wind velocity. The value $\epsilon_{\rm
  SN} \approx 1.8\times 10^{49}~\erg\,\Msun^{-1}$ 
is appropriate for a \cite{Chabrier2003} initial mass
function (IMF) and a stellar mass range 0.1--100~$\Msun$ if all stars 
in the mass range 6--100$~\Msun$ end their lives as core-collapse SNe. For
comparison, assuming a \cite{Salpeter1955} IMF would yield $\epsilon_{\rm SN}
\approx 1.1\times 10^{49}~\erg\,\Msun^{-1}$. We will use $\eta = 2$
and $v_{\rm 
  w} = 600~\kms$ as our fiducial values, which implies that the winds
carry about 
forty percent of the energy produced by core collapse SNe. The
remainder is implicitly assumed to be lost radiatively. 

These
values are consistent with observations of local starburst galaxies,
for which $v_{\rm w}$ increases with galaxy mass (or star formation
rate) from tens to $\sim 10^3~\kms$, while the 
mass in cold, outflowing gas ranges from $\eta \sim 0.01$ to 10
\cite[e.g.][and references therein]{Veilleux2005}, and with redshift $z\sim
3$ starburst galaxies, which typically have wind velocities of
hundreds to $\sim 10^3~\kms$ \cite[e.g.][]{Shapley2003}. Comparisons with
observed values are, however, plagued by the fact that they apply
only to certain gas phases and that it is
not clear on what scale they are measured. Moreover, it is not clear a
priori how the wind mass loading and velocity predicted by the
simulation correspond to the input values. In fact, studying this question is
one of the main motivations for the present work.

Our recipe for kinetic feedback in SPH simulations is as follows. 
Once a star particle reaches an age $t_{\rm SN} = 3\times 10^7~\yr$,
corresponding to the maximum lifetime of stars that end their lives as
core-collapse SNe, it is allowed to inject
kinetic energy into its surroundings by kicking one or more of its
neighbours. 
We will refer to particles that have received a kick in the
time interval $[t-t_{\rm w},t]$, where $t$ is the current time, as wind
particles. New wind particles are selected 
stochastically from the neighbouring gas particles of each newly-formed
star particle.\footnote{We ensured that the result is independent of
  the order in which the particles are processed. The probability that
  a given gas particle becomes a wind particle is evaluated using the
  sum of the unique particle IDs of the gas and the neighboring newly
  formed star particle as the seed for a random number generator. The
  latter uses a table of random numbers that is updated at every time
  step. A gas particle becomes a wind particle if it is selected by at
  least one newly formed star particle. The direction in which the
  particle is kicked is randomized using the particle ID as the seed
  and is therefore also independent of the order.}
 That is, each neighbouring gas particle has a probability
\begin{equation}
\label{eqn:wprob}
Prob =\eta\frac{m_{\ast}}{\Sigma_{i=1}^{N_{\rm ngb}} m_{g,i}}
\end{equation}
of becoming a wind particle, where $m_{\ast}$ is the mass
of the new star particle, $m_{{\rm g},i}$ the mass of gas particle $i$,
$N_{\rm ngb}$ the number of neighbours 
within the SPH smoothing kernel ($N_{\rm ngb}=48$ in our simulations),
and the sum is over all gas particle neighbours that are not wind
particles. The wind velocity is then added to the velocity of the wind
particle with a random direction.\footnote{We implemented the
  possibility of giving the wind velocity the direction of the vector
  pointing from the star to the wind particle, but found no noticeable
  effects. Contrary to randomized directions, for directed kicks the
  result is not independent of the order in which the particles are
  processed because it matters which newly formed star particle
  selected the wind particle first.}

The small time delay $t_{\rm SN}$ is unimportant for the simulations
presented here, but we implemented it because it is physically
sensible: it takes $t_{\rm SN}$ for the feedback energy to be
released. We note that it may be important for simulations that follow
the ejection of heavy elements by massive stars as a function of time
because it prevents gas particles from being kicked before they are
enriched.

The parameter $t_{\rm w}$ determines how long a
gas particle that is kicked will remain classified as a wind
particle. We do not allow wind particles to be kicked. This ensures
that the wind velocity does not become much greater than $v_w$,
which would not only complicate the interpretation of the simulations,
but could also result in extremely small time steps. We also do not
allow wind particles to be converted to star particles, in order to
prevent the formation of spurious high velocity star particles. We set
$t_{\rm w} = 1.5\times 10^7~\yr$, but note that similar or smaller
values (including $t_{\rm w} =0$) give nearly identical results. 

It is clear that for $\eta>N_{\rm ngb}$, and potentially for lower
values of $\eta$ in regions of intense star formation, and/or for large
values of $t_{\rm w}$, it can happen
that all the gaseous neighbours of a newly-formed star particle are
wind particles. For our default parameter values this problem is, in
fact, negligible. However, to avoid a break down of the method for
extreme parameter values, we keep track of the desired and actual
global, cumulative mass in wind particles. If the
actual mass falls below the desired mass (i.e.\ below $\eta M_\ast$,
where $M_\ast$ is the total stellar mass in the simulation),
additional wind
particles are selected stochastically from all 
star-forming particles that are not currently wind particles. Thus, this
correction to the mass loading is a non-local form of kinetic
feedback, much as is done for all wind particles in SH03. We stress the
fact that this correction
to the mass loading is negligible (i.e.\ $\la 1\%$) for all
simulations presented here.

\begin{figure*}
\includegraphics[width=0.327\textwidth]{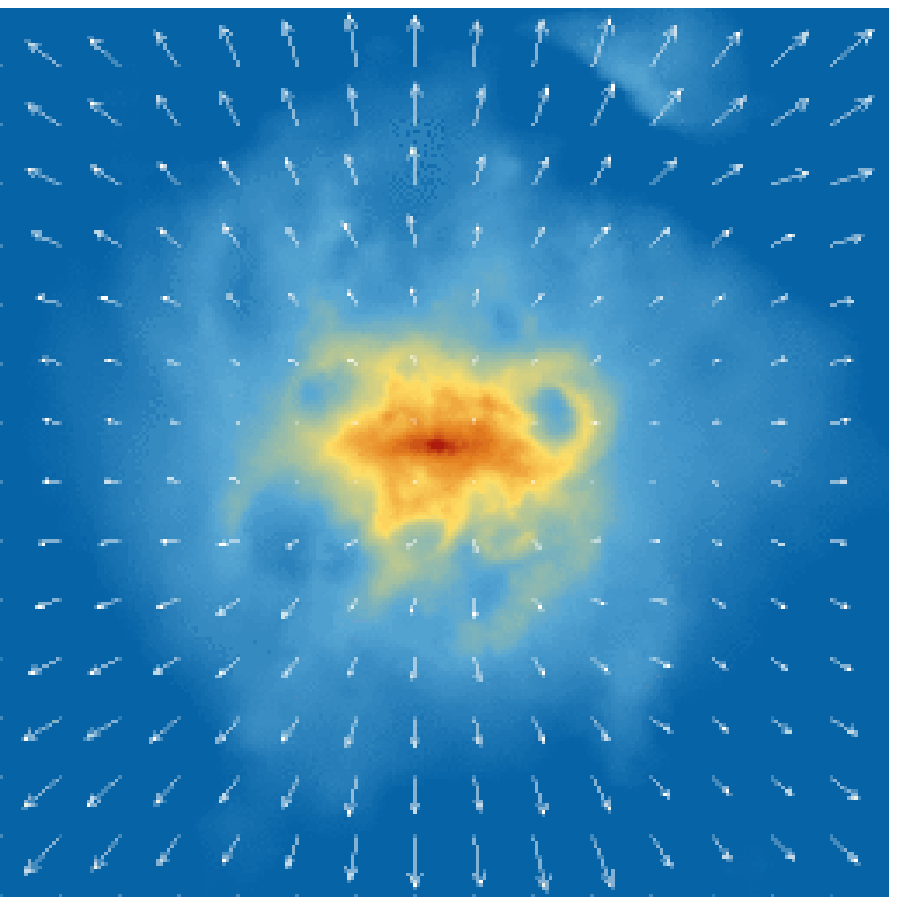}%
\includegraphics[width=0.327\textwidth]{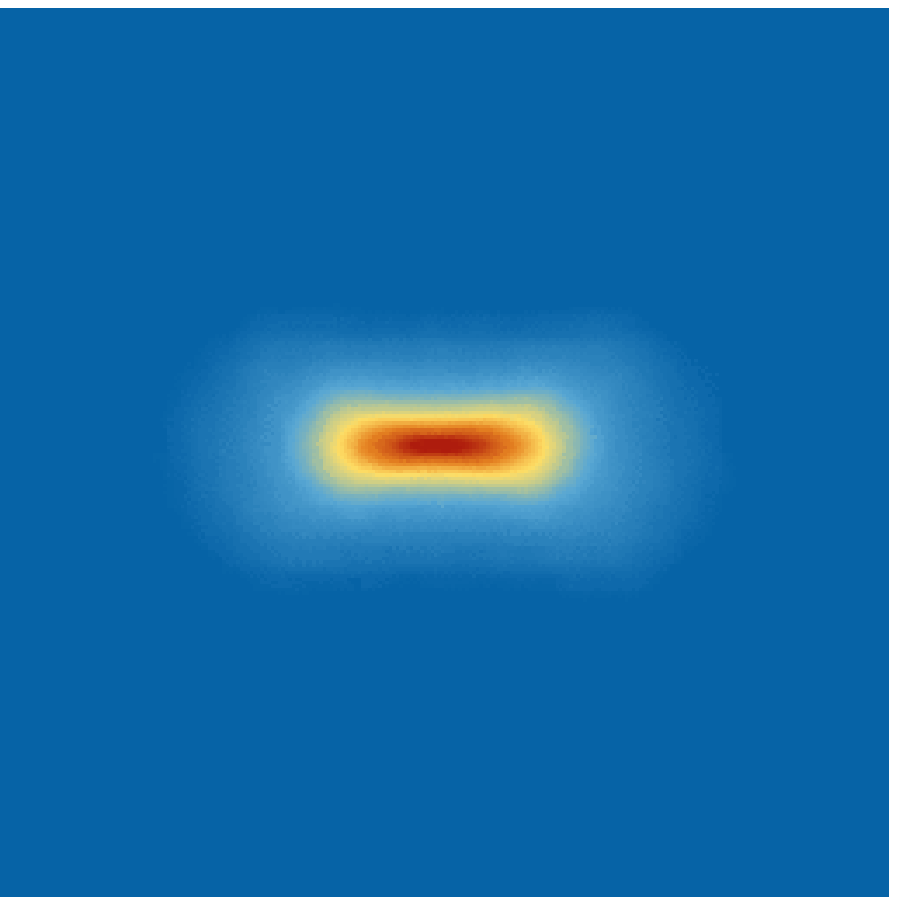}%
\includegraphics[width=0.327\textwidth]{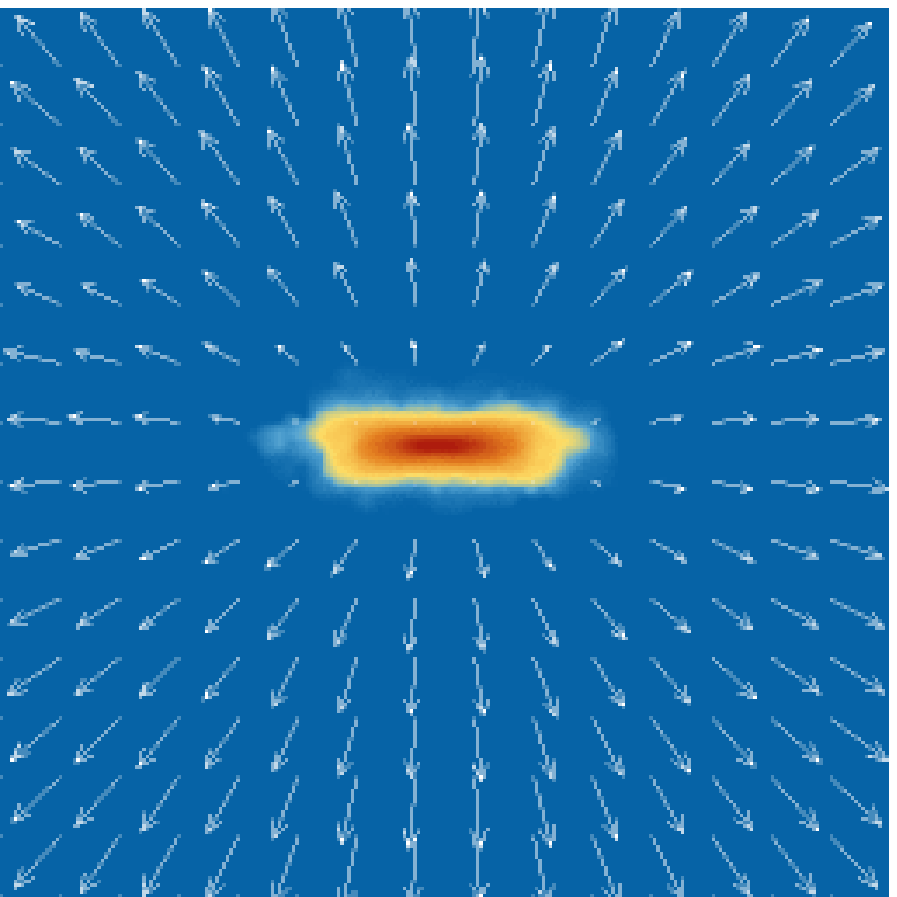}\\
\includegraphics[width=0.327\textwidth]{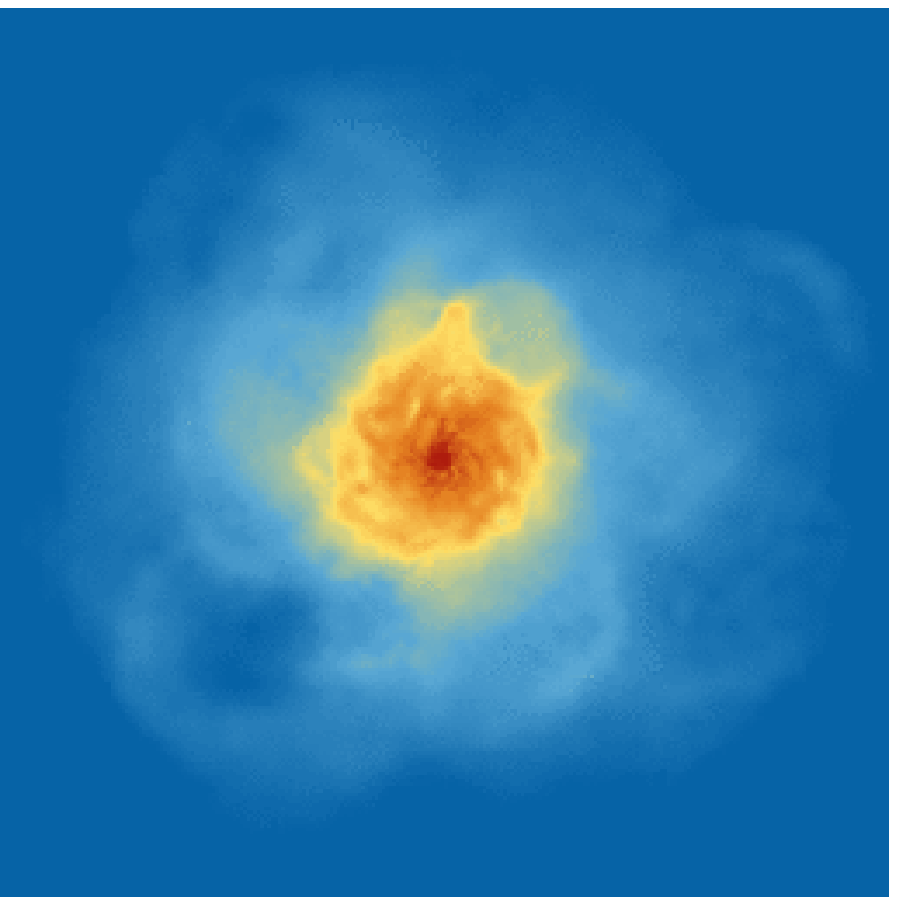}%
\includegraphics[width=0.327\textwidth]{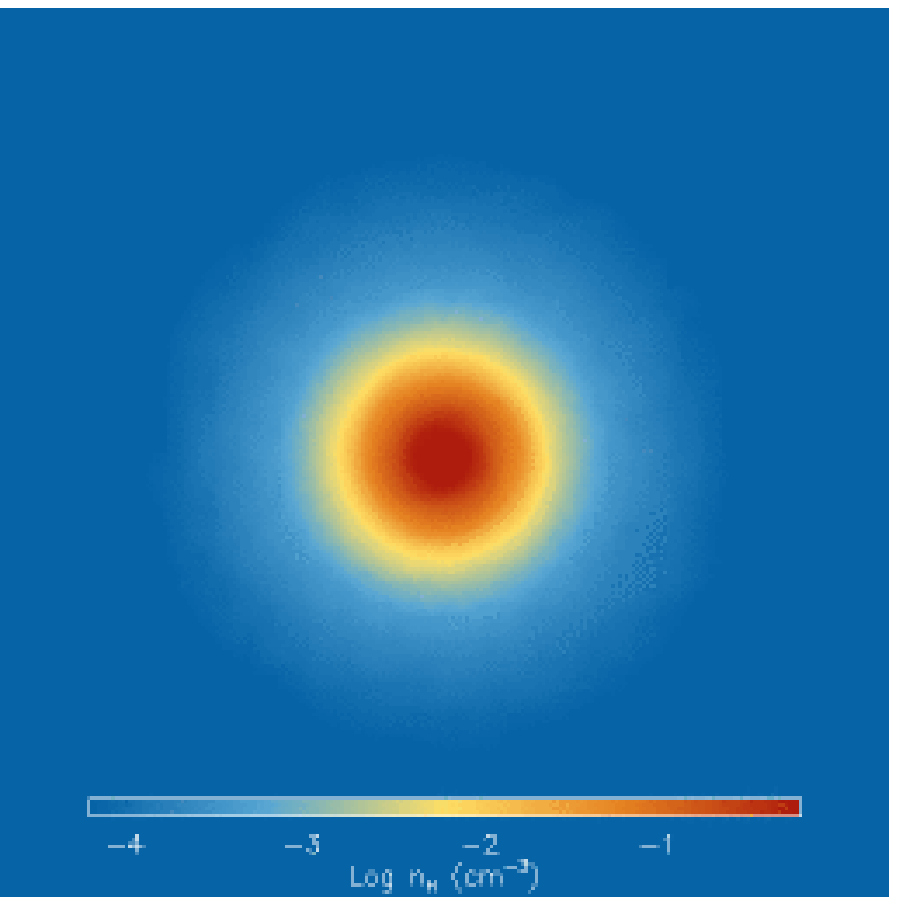}%
\includegraphics[width=0.327\textwidth]{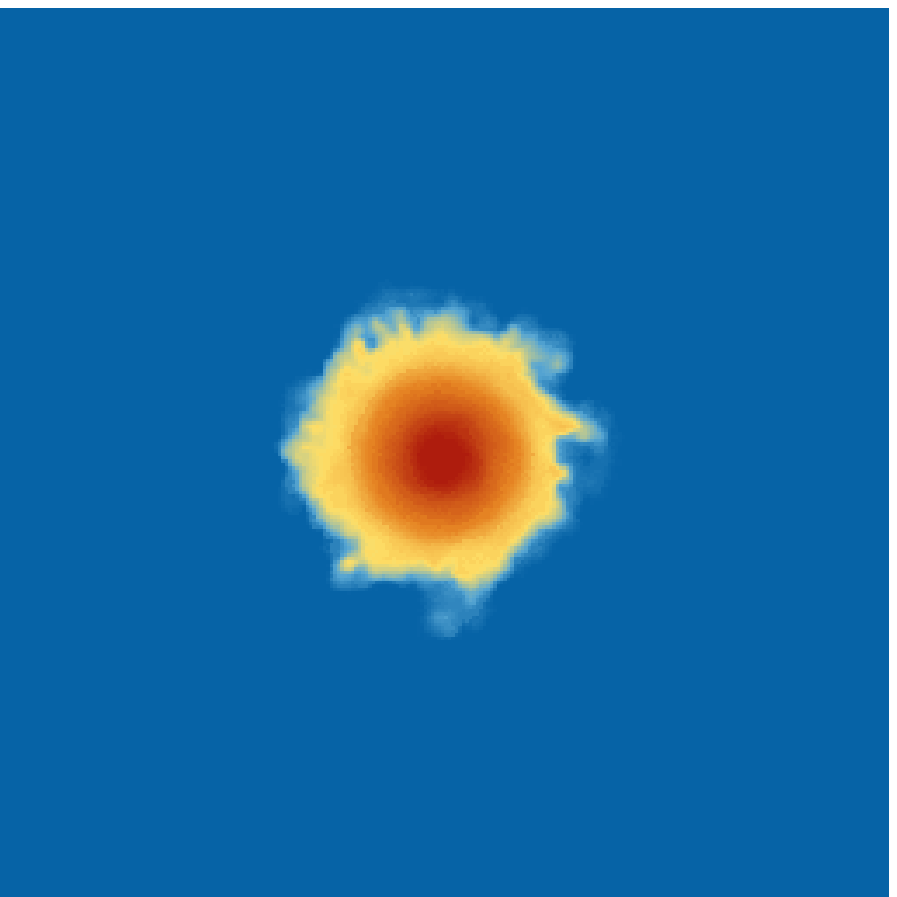}%
\caption{Edge- and face-on projections of the gas density
  for models \textit{m10} (left-hand column), \textit{m10nowind}
  (middle column), and \textit{m10dec} (right-hand column) at time
  $t=250~\Myr$. While the default galactic wind dramatically alters
  the morphology, the model using decoupled winds resembles
  the model without winds. Images are $17.5~\kpch$ on a side and only
  show the gas component of the disc. The color coding is logarithmic
  in density. The color scale is fixed in each image and is indicated by
  the color bar in the lower middle panel. The maximum vector length
  corresponds to a velocity of $106~\kms$ for \textit{m10} and
  $502~\kms$ for \textit{m10dec}. The velocity field is not shown for
  model \textit{m10nowind}.}
\label{fig:m10}
\end{figure*}

\section{Simulations}
\label{sec:sims}

We ran simulations of isolated disc galaxies embedded in dark matter
halos with total masses of $10^{10}$ and $10^{12}~\Msolh$ (we use
$h=0.73$). The models do not include gaseous halos. Initially all the
gas is therefore in the discs. Runs were repeated using 
varying physical and numerical parameters. In this section we
will briefly describe the code, the initial conditions and the runs
we performed. We note that the code and initial conditions are
identical to those used in \cite{Schaye2008}.

\subsection{Code and initial conditions}
\label{sec:code}

We use a modified version of the TreePM/SPH code \textsc{gadget}
\citep{Springel2001,Springel2005} for all the simulations presented in
this paper. 

We employ the star formation recipe of \cite{Schaye2008}, to which we
refer the reader for details. Briefly, 
gas with densities exceeding the critical density for the
onset of the thermo-gravitational instability ($n_{\rm H} \sim 10^{-2}
- 10^{-1}~\cm^{-3}$) is
expected to be multiphase and star-forming
\citep{Schaye2004}. We therefore impose an
effective equation of state with pressure $P\propto \rho_g^{\gamma_{\rm
    eff}}$ for densities exceeding $n_{\rm H} =
0.1~\cm^{-3}$, normalized to $P/k = 10^3~\cm^{-3}\,\K$ at the
threshold. We use $\gamma_{\rm eff} = 
4/3$  for which both the Jeans mass and the ratio 
of the Jeans length and the SPH kernel are independent of the density,
thus preventing spurious fragmentation due to a lack of numerical resolution. 

The Kennicutt-Schmidt star formation law is analytically converted
and implemented as a pressure law.  As we
demonstrated in \cite{Schaye2008}, our method
allows us to reproduce arbitrary input star formation laws for any
equation of state without tuning any parameters. All the simulations
presented here produce similar local Kennicutt-Schmidt laws, which
implies that comparisons of the predicted star formation laws with
observations cannot discriminate between different wind models. We use 
the observed 
\cite{Kennicutt1998} law 
\begin{equation}
\dot{\Sigma}_\ast = 1.5 \times 10^{-4} \,\Msolyrkpcsq \left
    ({\Sigma_g 
    \over 1~\Msolpcsq}\right )^{1.4},
\label{eq:KS}
\end{equation}
where we divided Kennicutt's normalization by a factor 1.65 to account for the
fact that it assumes a Salpeter IMF whereas we are using a Chabrier
IMF.\footnote{Note that we assumed a Salpeter IMF in
  \cite{Schaye2008}.}

Radiative
cooling and heating were included using tables for hydrogen and helium,
assuming ionization equilibrium in the presence of the
\cite{Haardt&Madau2001} model for the $z=0$ UV background radiation
from quasars and galaxies. The cooling tables were generated using the
publicly available package \textsc{cloudy} \citep[version
  06.02;][]{Ferland2000}. 

The initial conditions
were generated with a code kindly provided to us by Volker
Springel. Detailed descriptions of the model and its implementation
are given in \cite{Springeletal2005}, so we will provide only a brief
summary here.

The model consists of a dark matter halo, a
stellar bulge, and an exponential disc of stars and gas. The circular
velocities at the virial radii are $35.1$ and $163~\kms$ for the
$10^{10}$ and $10^{12}~\Msolh$ halos, respectively. The dark matter
halo and the stellar bulge both follow \cite{Hernquist1990} profiles
scaled to match the inner density profile of a \cite{NFW} (hereafter NFW) profile with 
concentration $c=9$ and a mass within the virial radius equal to the
total mass of the Hernquist profile. The equivalent NFW virial radii
are then $35.1$ and $163~\kpch$ for the $10^{10}$ and $10^{12}~\Msolh$
halos, respectively.  The halo has a dimensionless spin parameter
$\lambda=0.33$.  The disc contains 4 percent of both the total mass
and the total angular momentum. The bulge contains 1.4 percent of the
total mass and has a scale length one tenth of that of the disc. The
bulge has no net rotation, while the dark matter halo and disc have the
same specific angular momentum. The initial gas fraction of the disc
is 30 percent, the remaining 70 percent of the disc mass is
made up of stars.  The vertical distribution of the
stellar disc follows the profile 
of an isothermal sheet with a constant scale height set to 10~percent of
the radial disc scale length. The vertical gas
distribution is set up in hydrostatic equilibrium using an iterative
procedure. 

\begin{figure}
\includegraphics[width=0.245\textwidth]{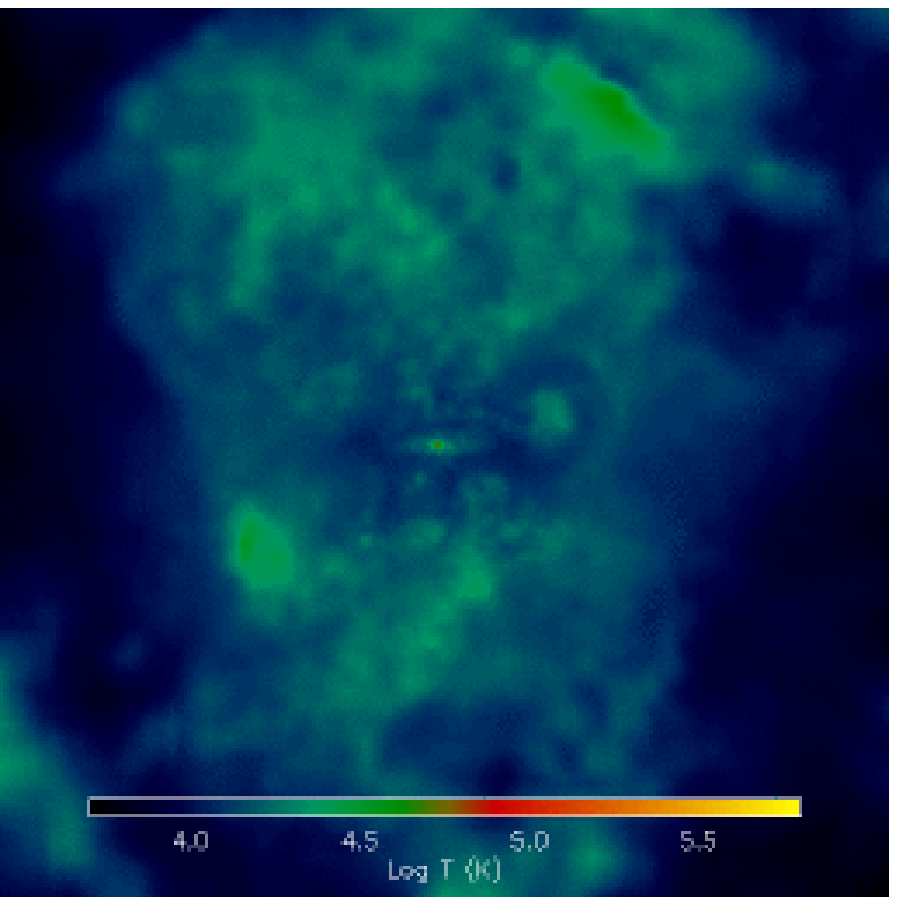}%
\includegraphics[width=0.245\textwidth]{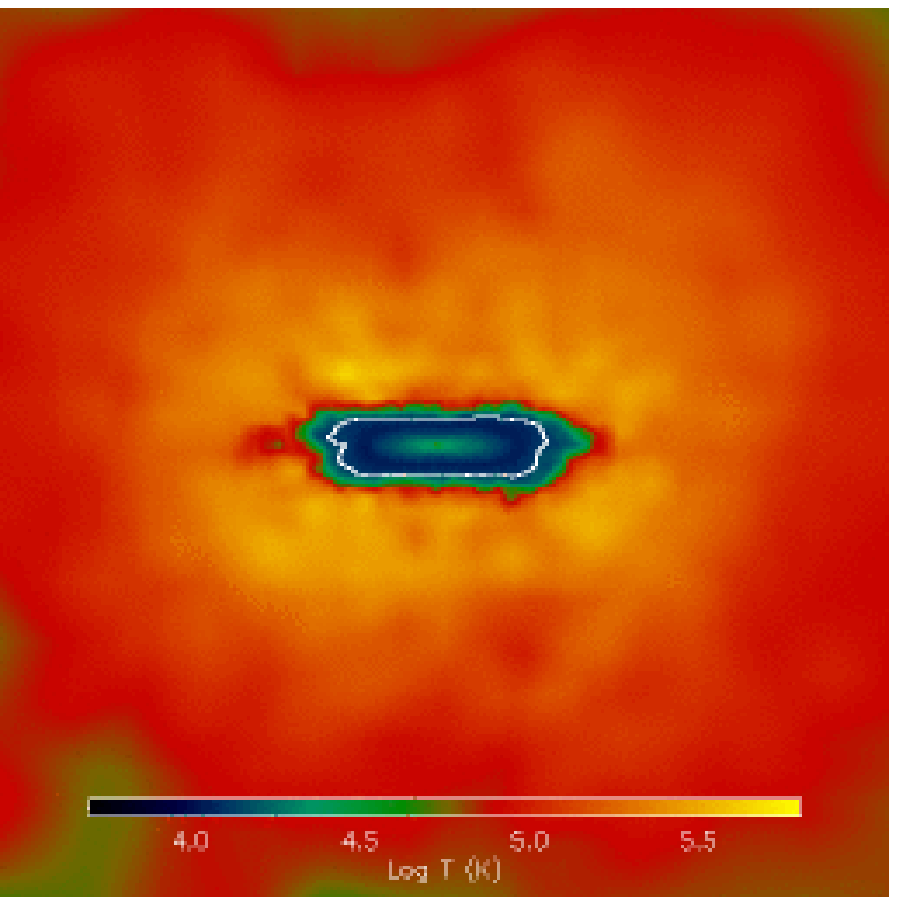}\\
\includegraphics[width=0.245\textwidth]{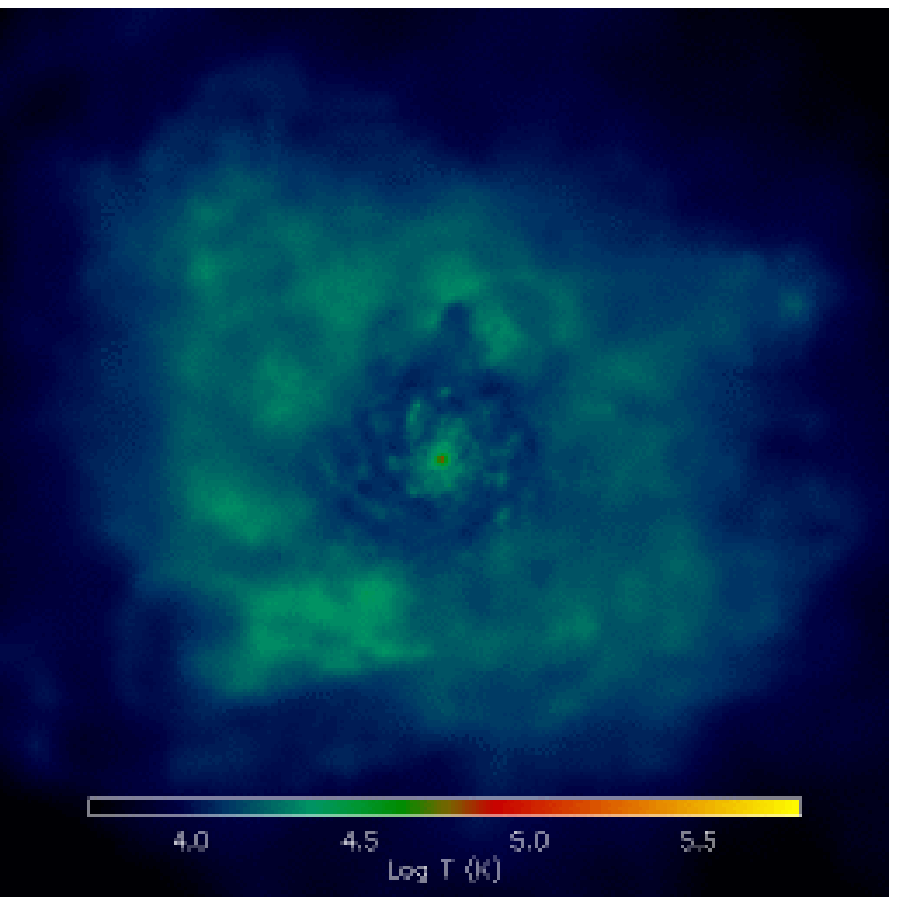}%
\includegraphics[width=0.245\textwidth]{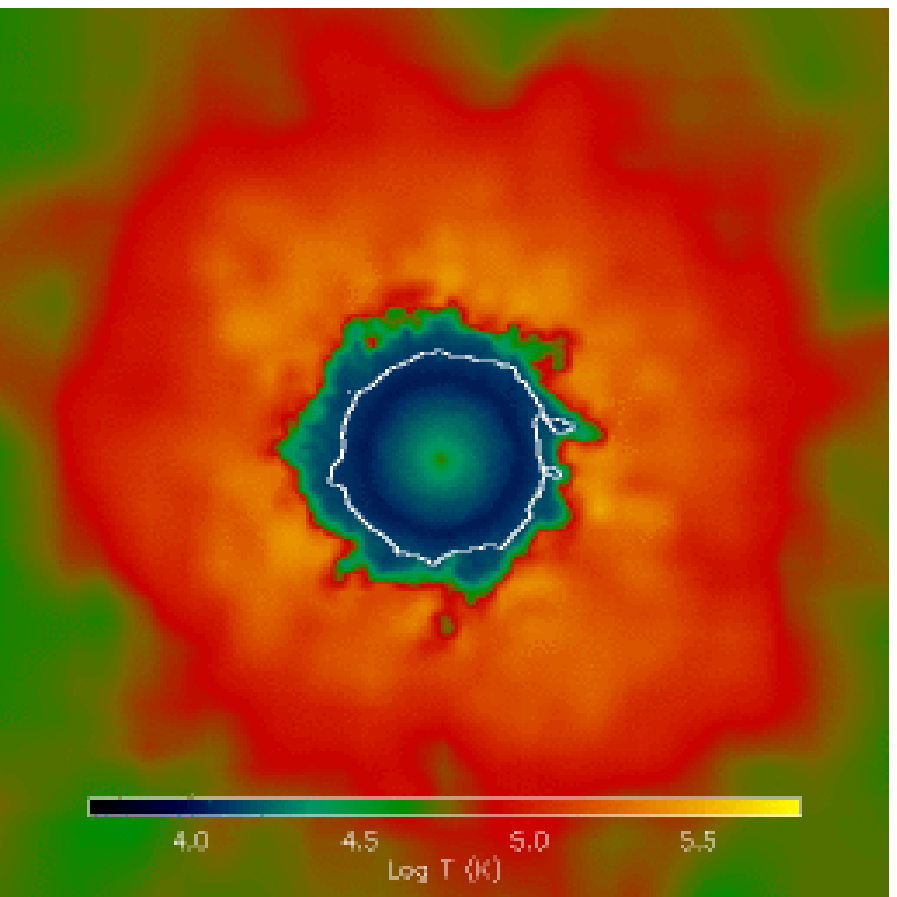}
\caption{Edge- and face-on projections of the gas temperature for
  models \textit{m10} (first column) and \textit{m10dec} (second
  column) at time $t=250~\Myr$. The white contours correspond to a gas
  density of $n_{\rm H} = 10^{-2}~\cm^{-3}$, which is the density
  below which the winds are recoupled to the hydrodynamics in the case
  of \textit{m10dec}. Images are $17.5~\kpch$ on a side. The color
  coding is logarithmic in temperature. The color scale is indicated
  by the color bar in each panel.}
\label{fig:m10temp}
\end{figure}

Except for our low-resolution runs, the  
total number of particles in each simulation is 4,800,736, of which
225,437 are gas particles in the disc. The baryonic particle mass for the
$10^{12}~\Msolh$ halo is $m_{\rm b}=5.1\times
10^4~\Msolh$. As discussed in \cite{Schaye2008}, this particle mass
implies, for star-forming gas, a constant ratio of the
total mass within a smoothing kernel to the Jeans mass, $N_{\rm
  ngb}m_{\rm b}/M_{\rm J}=1/6$, and a constant ratio of the kernel to
the Jeans length $h/L_{\rm J} = 1/(48)^{1/3}$. Although the $10^{10}~\Msolh$ halo
requires a factor 100 less particles to resolve the Jeans scales by
the same margin, we choose to use the same number of
particles as for the $10^{12}~\Msolh$ halo in order to achieve a similar
sampling of the outflows. The dark matter
particle mass is higher than that for baryonic particles by a factor
$(\Omega_m-\Omega_b)/\Omega_b\approx 4.6$. 

\begin{figure}
\includegraphics[width=0.245\textwidth]{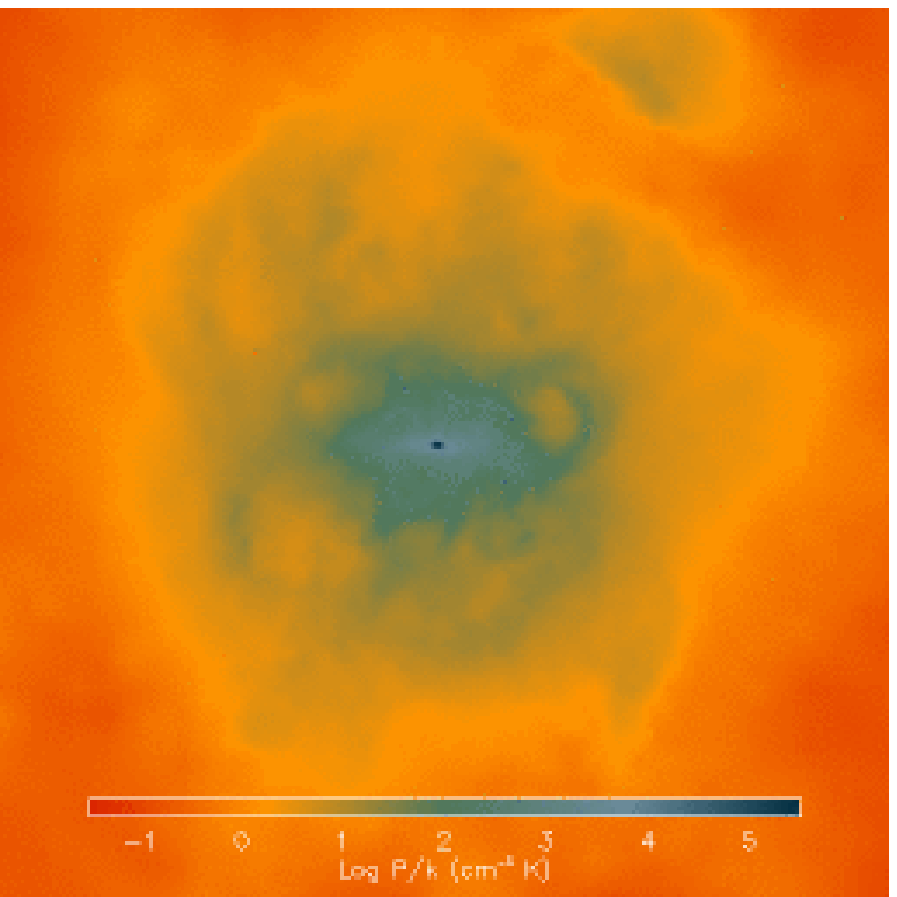}%
\includegraphics[width=0.245\textwidth]{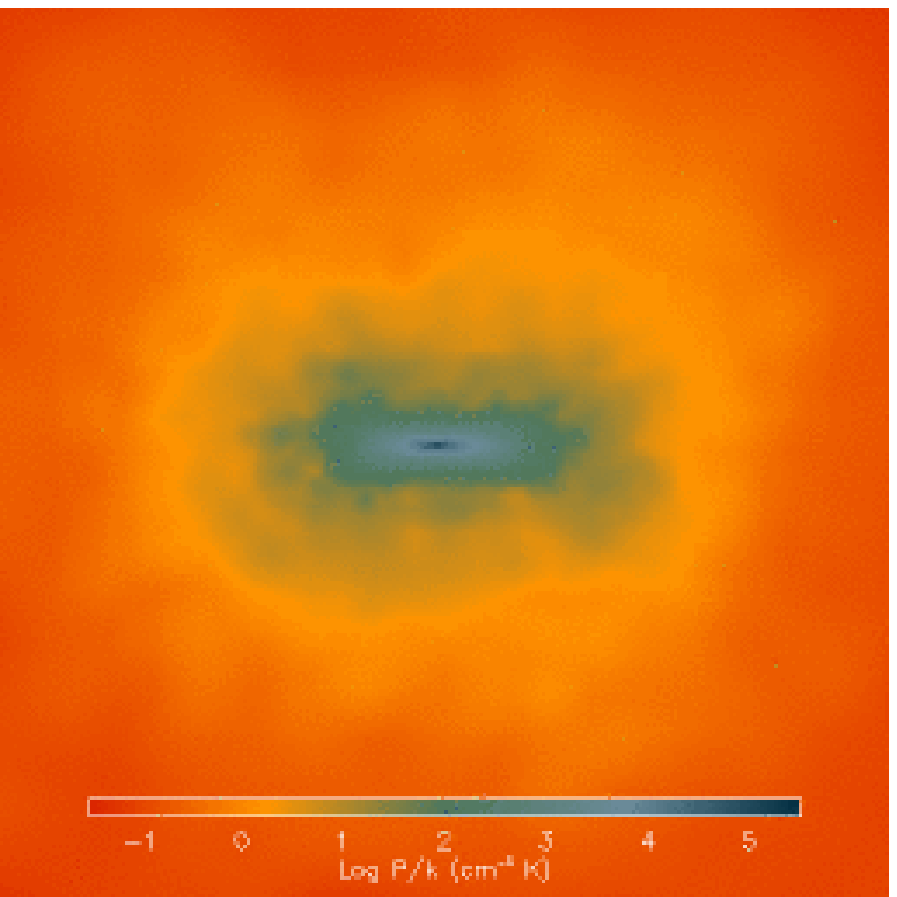}\\
\includegraphics[width=0.245\textwidth]{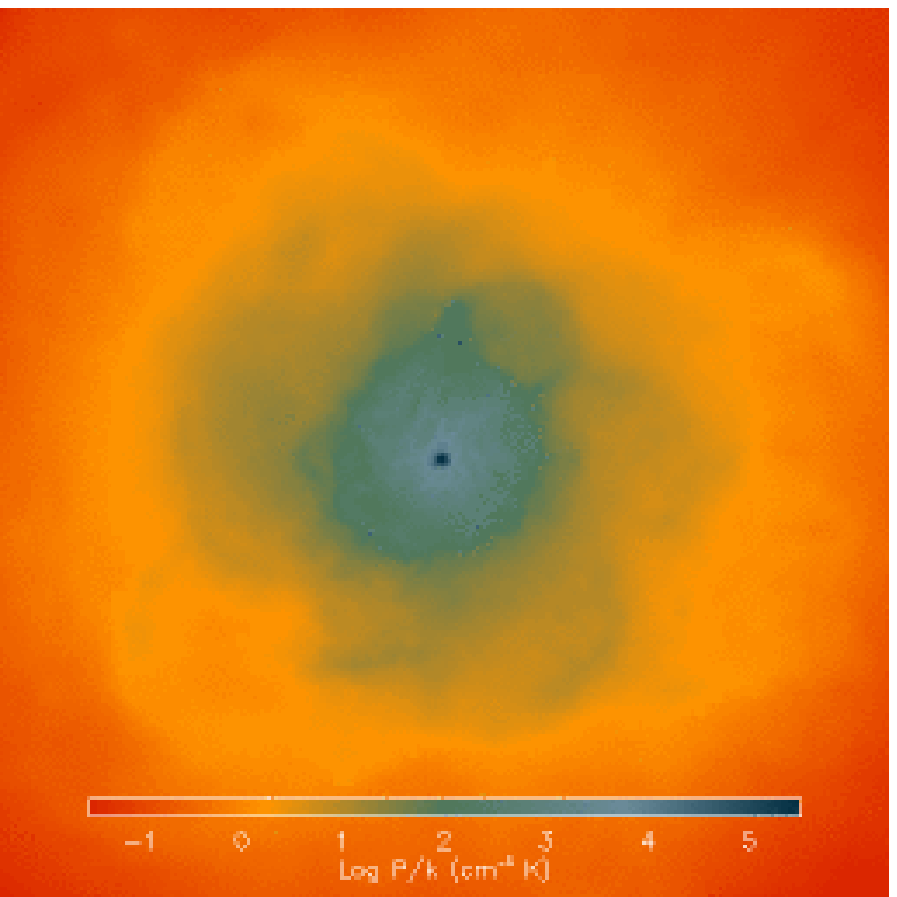}%
\includegraphics[width=0.245\textwidth]{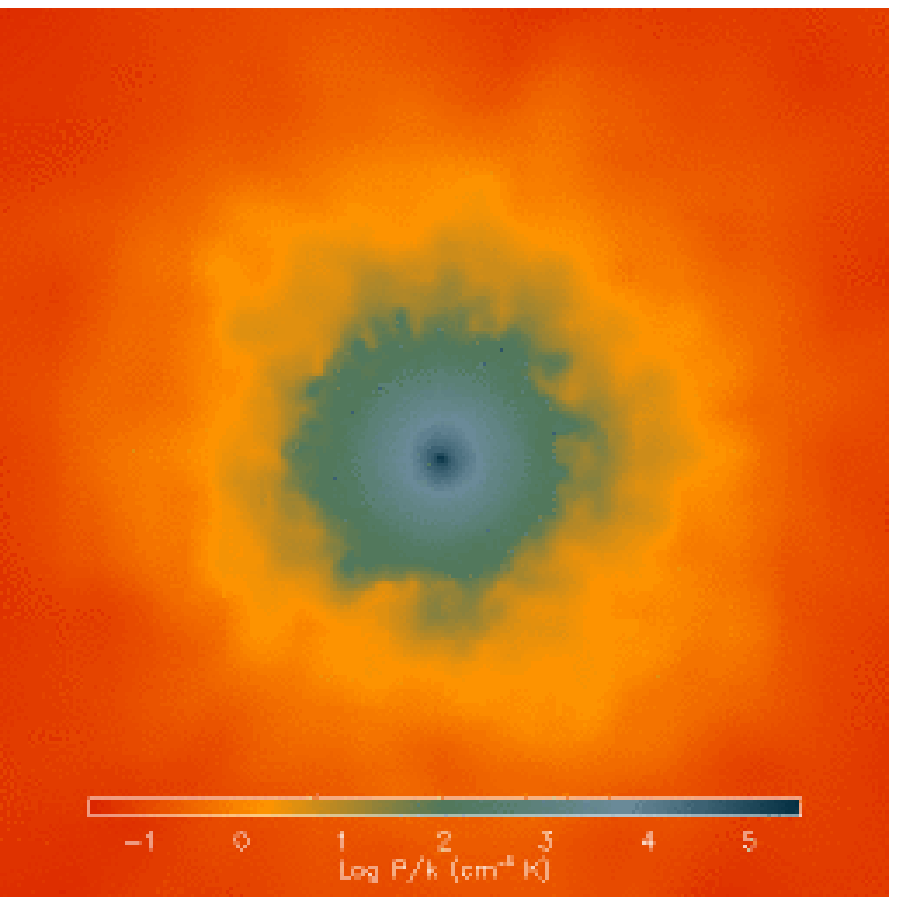}
\caption{Edge- and face-on projections of the gas pressure for models
  \textit{m10} (first column) and \textit{m10dec} (second column) at
  time $t=250~\Myr$. Images are $17.5~\kpch$ on a side. The color
  coding is logarithmic in pressure. The color scale is indicated by
  the color bar in each panel.}
\label{fig:m10pressure}
\end{figure}

The gravitational softening length was set to $\epsilon_b=10~\pch$ for
 the baryons and to $(m_{\rm 
 dm}/m_b)^{1/3}\epsilon_b\approx 17~\pch$ for the dark matter. This is
sufficiently small to resolve the Jeans length by at least two
softening lengths up to gas surface densities of $\sim
10^{4.5-5}~\Msolpcsq$, which greatly exceed the central surface
density for both halos.

\subsection{Runs}

Table~\ref{tbl:params} lists all 16 simulation runs we have
performed. Each simulation was evolved for $500~\Myr$. Our fiducial
simulations are labelled \textit{m10} and \textit{m12} for the $10^{10}$
and the $10^{12}~\Msolh$ halos, respectively. These runs used the
recipe for galactic winds described in section~\ref{sec:recipe}, with
mass loading $\eta=2$ and wind velocity $v_{\rm w}=600~\kms$. In
addition, we carried out the following variations:
\begin{itemize} 
\item One run without galactic winds (\textit{m[10,12]nowind}).
\item Two runs with the same input wind energy as the fiducial model,
  but with $(\eta,v_{\rm  w})=(1, 848~\kms)$ and $(4, 424~\kms)$,
  respectively (\textit{m[10,12]$\eta$1v848} and \textit{m[10,12]$\eta$4v424});
\item One run in which the wind particles are temporarily decoupled
  hydrodynamically (\textit{m[10,12]dec}); 
\item Three runs in which the number of particles was decreased by
  factors of 8, 64 and 512, respectively (\textit{m[10,12]lr008},
  \textit{m[10,12]lr064}, \textit{m[10,12]lr512})
\end{itemize}
In runs \textit{m[10,12]dec} the wind particles are decoupled from
the hydrodynamics in exactly the same manner as in SH03. That is, wind
particles do not feel and do not 
contribute to the pressure forces for a time $50~\Myr$
(corresponding to a distance of 31~kpc for a velocity of
$600~\kms$) or until their density has fallen below 10\% of the threshold
density for SF (i.e.\ until $n_{\rm H} < 10^{-2}~\cm^{-3}$), whichever
occurs first.

\begin{figure*}
\includegraphics[width=0.327\textwidth]{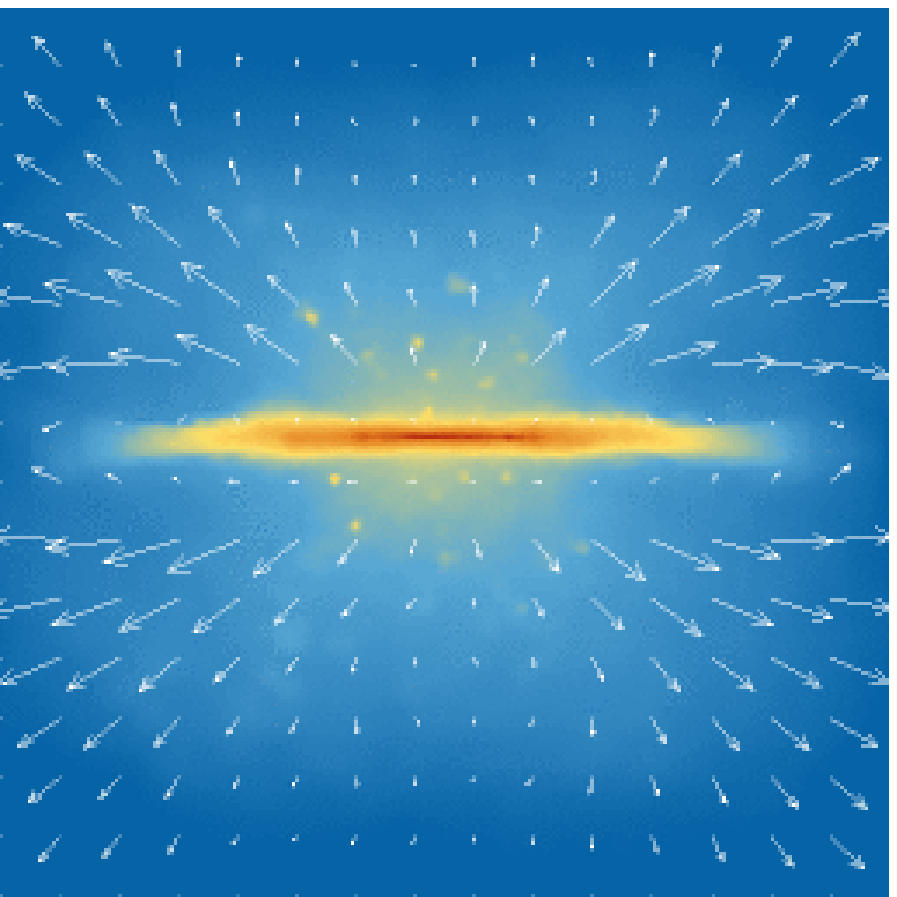}%
\includegraphics[width=0.327\textwidth]{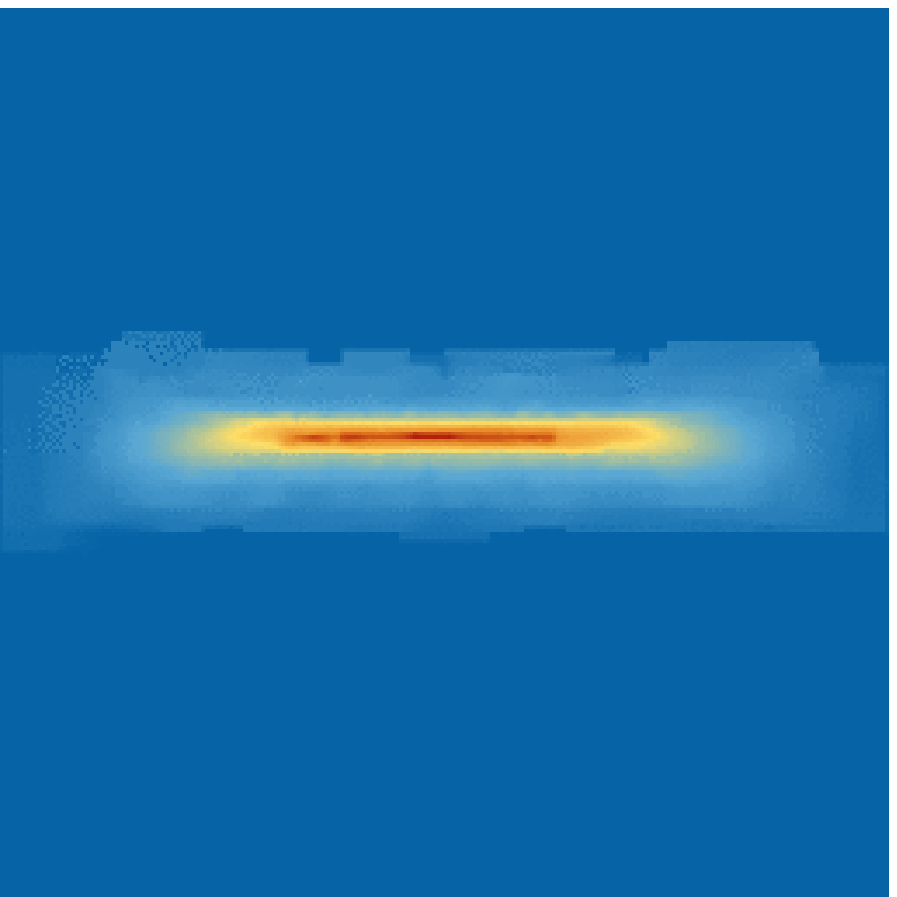}%
\includegraphics[width=0.327\textwidth]{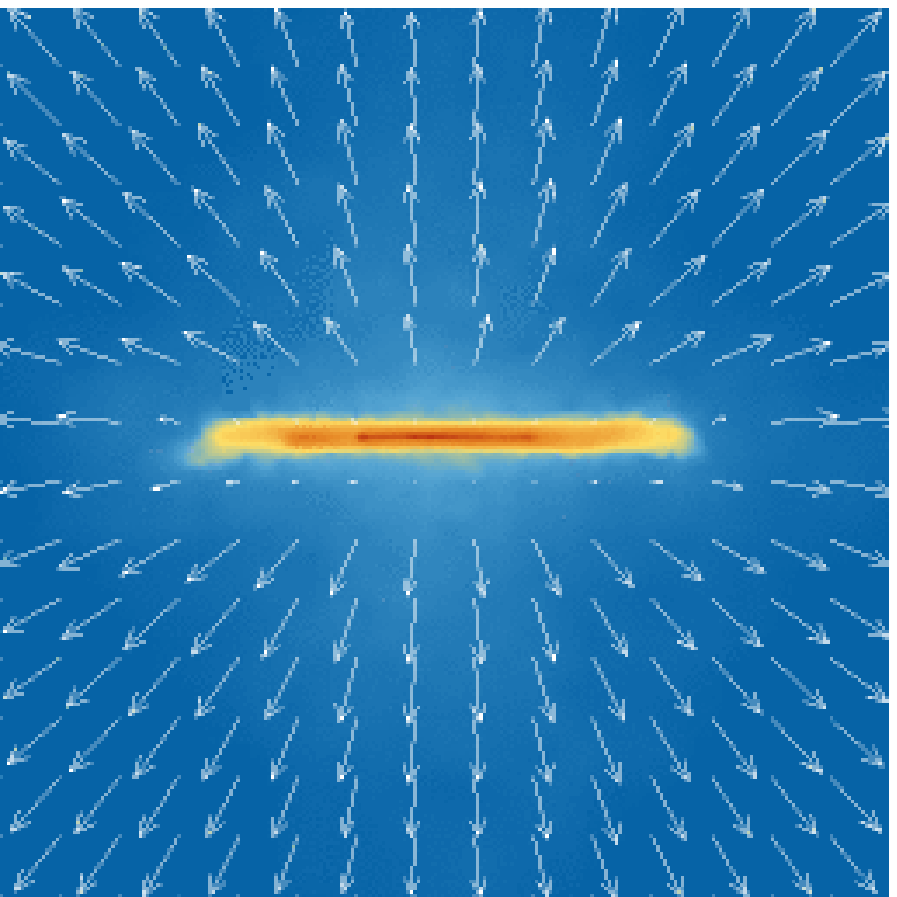}\\
\includegraphics[width=0.327\textwidth]{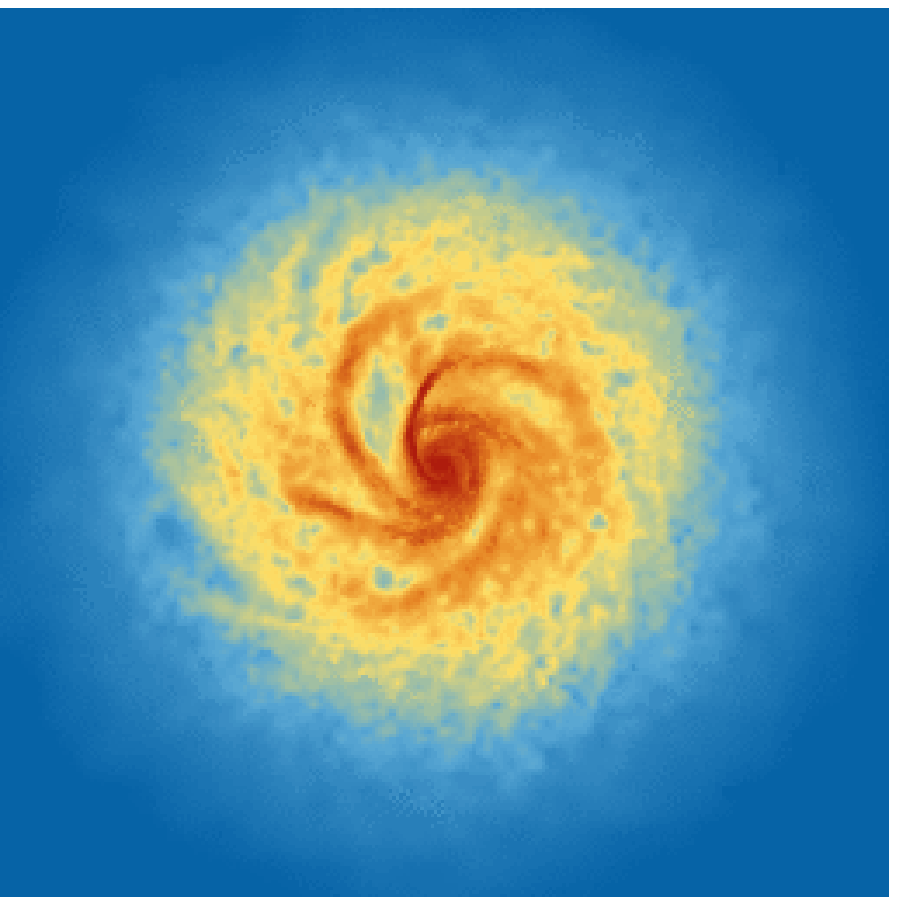}%
\includegraphics[width=0.327\textwidth]{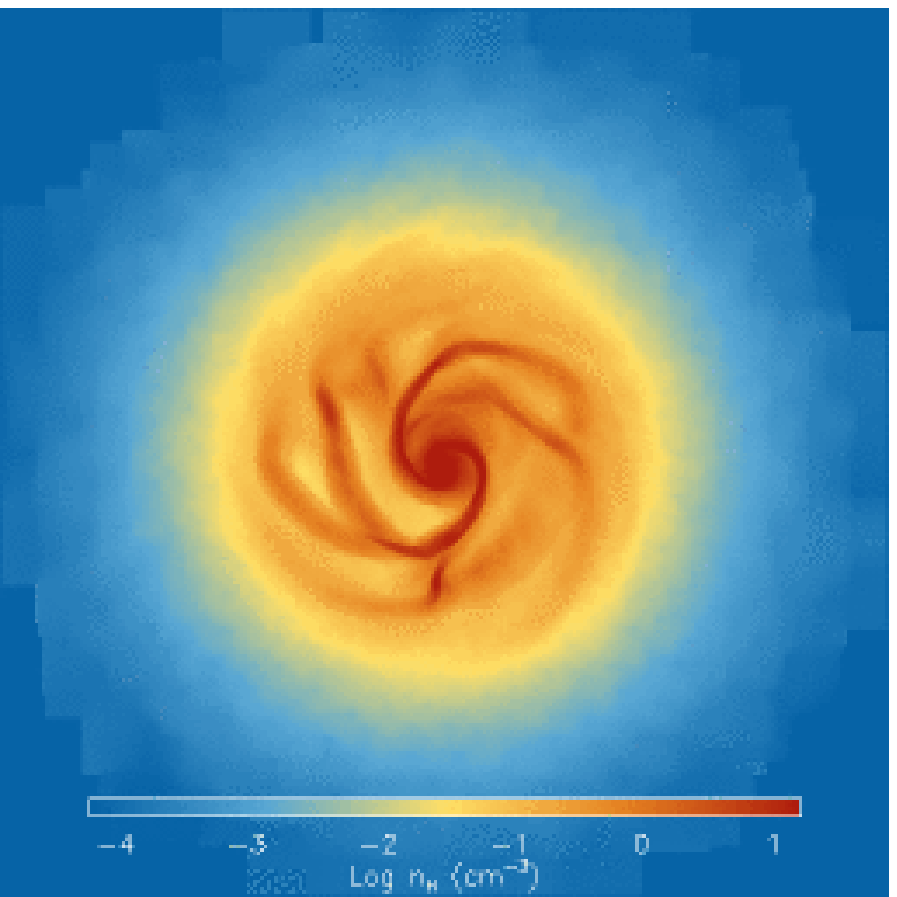}%
\includegraphics[width=0.327\textwidth]{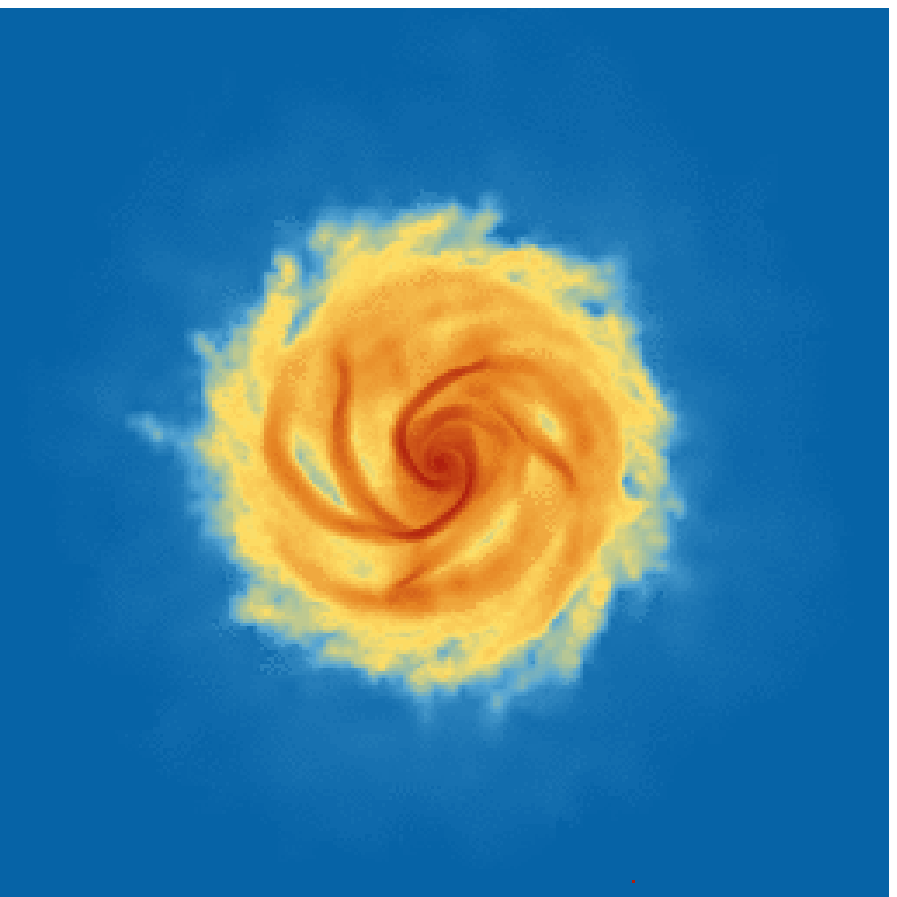}%
\caption{Edge- and face-on projections of the gas density
  for models \textit{m12} (left-hand column), \textit{m12nowind}
  (middle column), and \textit{m12dec} (right-hand column) at time
  $t=250~\Myr$. The default wind model initially produces a bi-conical
  outflow that develops into the galactic fountain shown here and it 
  results in a larger and more stable disc. Hydrodynamically
  decoupled winds produce a high-velocity wind, but have little impact
  on the morphology. Images are 
  $45~\kpch$ on a side and only show the gas 
  component of the disc. The color coding is logarithmic
  in density. The color scale is fixed in each image and is indicated by
  the color bar in the lower middle panel. The maximum vector length
  corresponds to a velocity of $141~\kms$ for \textit{m12} and
  $396~\kms$ for \textit{m12dec}.}
\label{fig:m12}
\end{figure*}

\section{Results}
\label{sec:res}

\subsection{Morphology}
\label{sec:morphology}

Fig.~\ref{fig:m10} shows edge- and face-on projections of the
gas density at time $t=250~\Myr$ for models \textit{m10} (left-hand column),
\textit{m10nowind} (middle column) and \textit{m10dec} (right-hand
column). A comparison of the first two columns shows that the
inclusion of a galactic wind has dramatic consequences for the
morphology of the dwarf galaxy. The wind blows low-density bubbles in
the disc, opening up channels through which it can escape into the
halo. The outflowing wind particles drag large amounts of disc gas
along, creating plumes of gas above and below the disc. The net result
is a substantially puffing up of the galaxy, giving it a diffuse and
irregular morphology, in good agreement with HI observations of nearby
dwarf galaxies
\cite[e.g.][]{Puche1992,Staveley-smith1997,Kim1999,Stanimirovic1999,Walter&Brinks1999}.  

The velocity field, which is indicated by the arrows in the edge on
projections, shows that our recipe for
galactic winds naturally produces a bi-conical outflow, with higher
velocities near the minor axis. The outflow velocities are typically a
few tens of km per sec, much lower than the input velocity of
$600~\kms$. The gas is therefore shocked to temperatures of only a few 
times $10^4~\K$ (Fig.~\ref{fig:m10temp}, left-hand column) and the pressure
closely tracks the gas density (Fig.~\ref{fig:m10pressure}, left-hand
column). 

In contrast to the fiducial model, the galaxy with hydrodynamically
decoupled winds (Fig.~\ref{fig:m10}, right-hand column) retains its initial 
morphology and the disc remains smooth and thin. Apart from a
small decrease of the density due to the removal of mass in the form
of the wind particles themselves, the only visible effect of the
decoupled winds is that the disc has become smaller and
thinner. Thus, in terms of morphology, the model 
with decoupled winds closely resembles the run
without galactic winds (middle column), except that the outer disc has
been removed. 

The decoupled winds decrease the size of the disc because gas at large
scale heights and/or large radii has a density 
sufficiently low for the wind particles to be recoupled
hydrodynamically (recall that wind particles are recoupled when their
density has fallen below 10 percent of the star formation threshold). Wind
particles are therefore able to remove the lowest density gas in the disc
but cannot replace it by dragging gas up from the inner parts of the
galaxy because of the temporary decoupling. Because the outer most
parts of the disc always have a low density, we find that the disc
continues to shrink throughout the simulation. 

\begin{figure}
\includegraphics[width=0.245\textwidth]{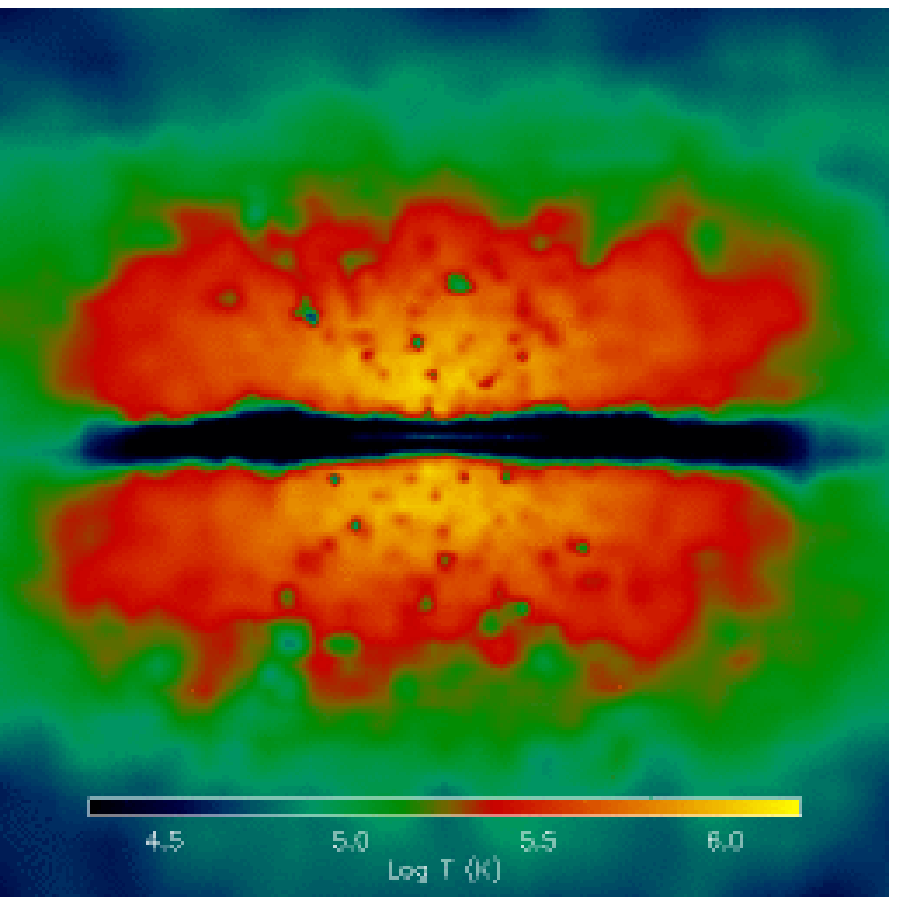}%
\includegraphics[width=0.245\textwidth]{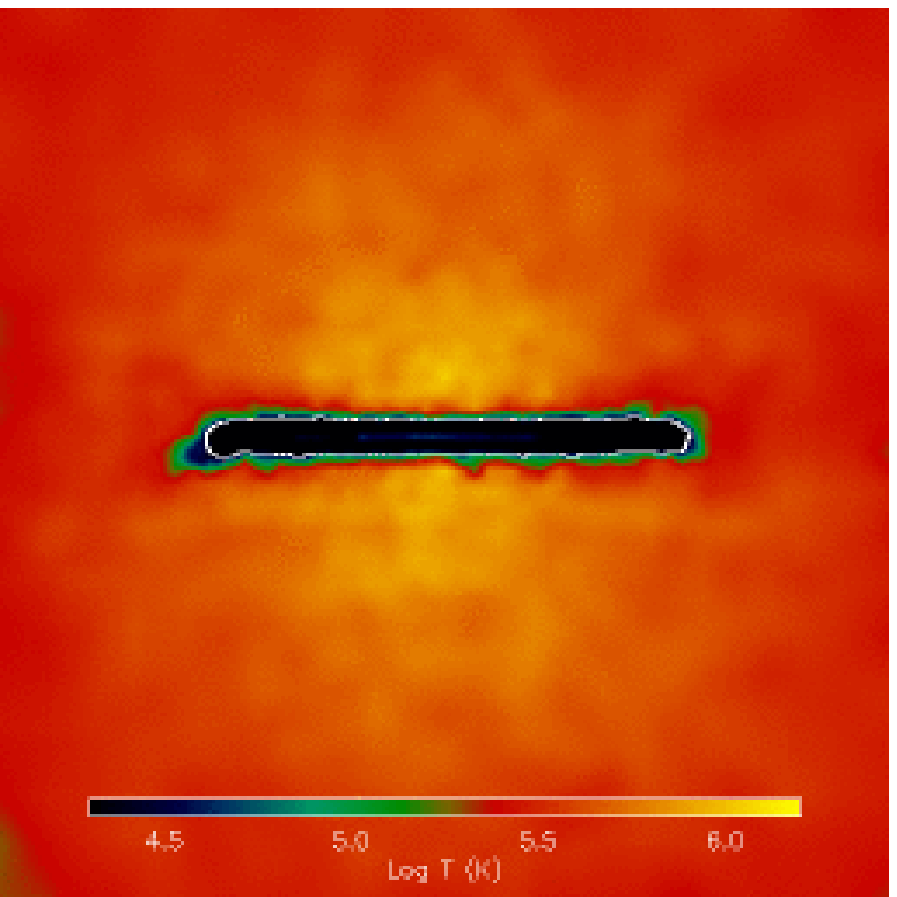}\\
\includegraphics[width=0.245\textwidth]{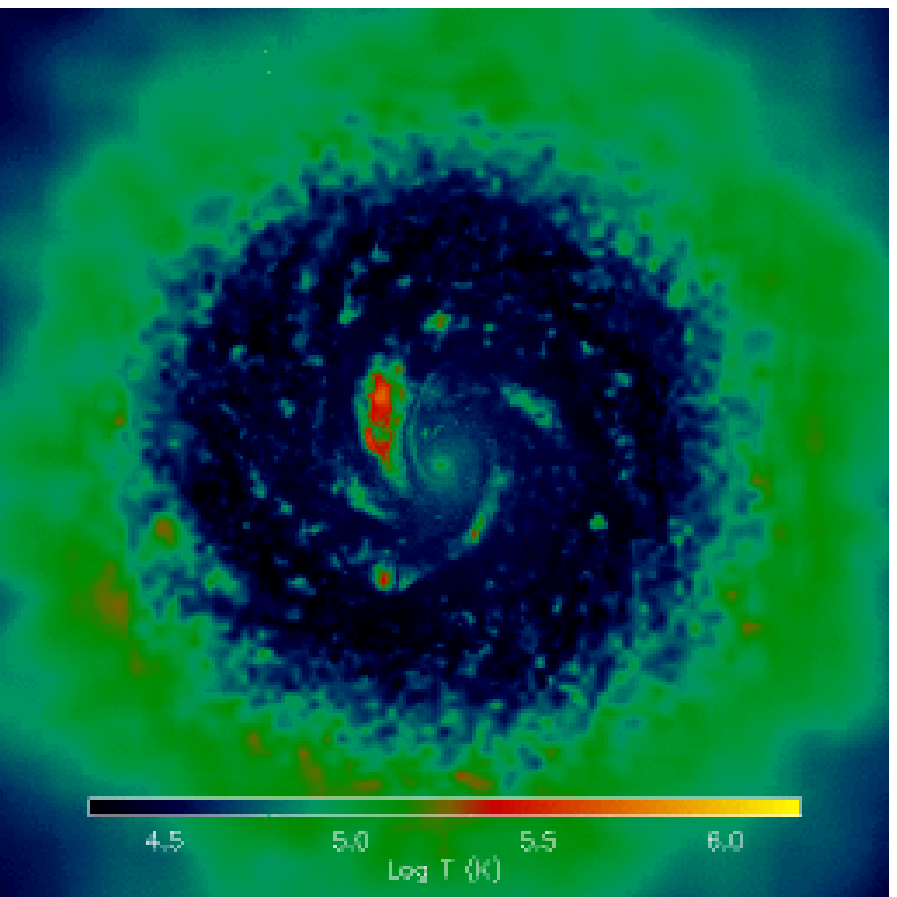}%
\includegraphics[width=0.245\textwidth]{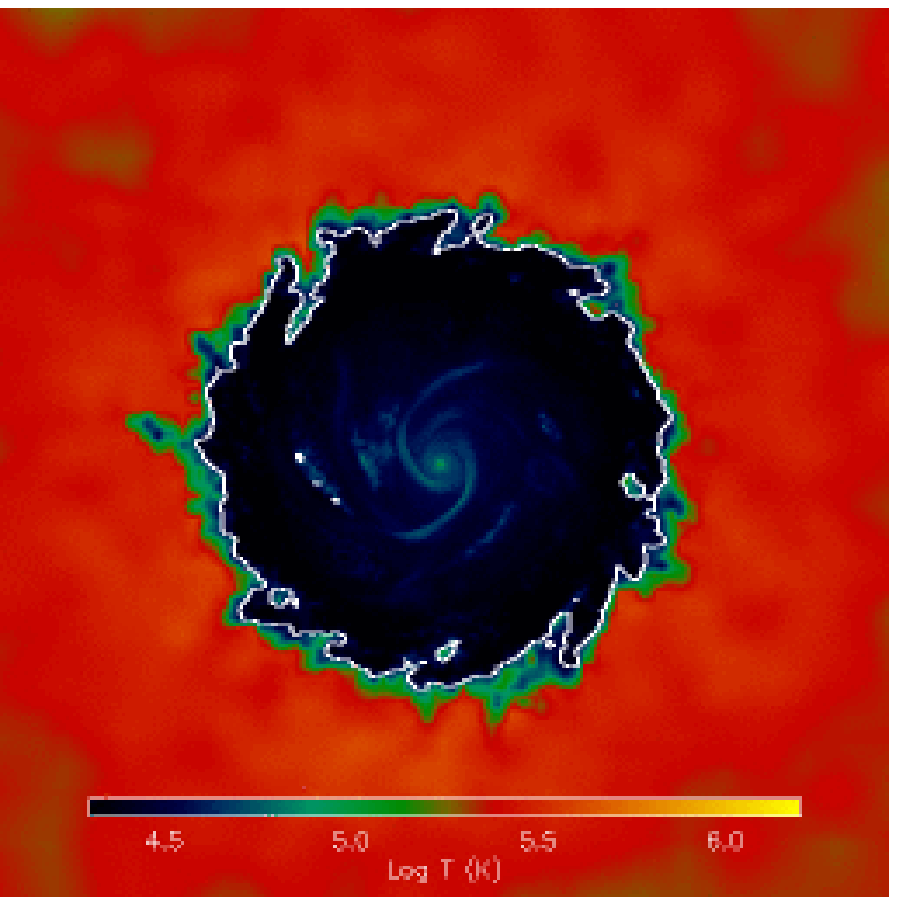}
\caption{Edge- and face-on projections of the gas temperature for
  models \textit{m12} (first column) and \textit{m12dec} (second
  column) at time $t=250~\Myr$. The white contours correspond to a gas
  density of $n_{\rm H} = 10^{-2}~\cm^{-3}$, which is the density
  below which the winds are recoupled to the hydrodynamics in the case
  of \textit{m12dec}. Images $45~\kpch$ on a side. The color coding
  is logarithmic in temperature. The color scale is indicated by the
  color bar in each panel.}
\label{fig:m12temp}
\end{figure}

The white contours in the right-hand columns of
Figures~\ref{fig:m10temp} and \ref{fig:m12temp} 
correspond to the density below which the wind particles are recoupled
to the 
hydrodynamics. It falls just inside the edge of the disc (by a margin
similar to the size of the local SPH kernel, which increases rapidly
with radius at the edge of the disc), confirming that
the size of the disc is set by the density below which the wind
particles are recoupled. As soon at the wind particles recouple, much
of their kinetic energy is converted into thermal energy through a
shock. Indeed, the figures show that the white contours 
closely track the local temperature minimum and that the edge
of the disc corresponds to a sharp jump in the temperature. The
temperature of the gas outside of the disc is much higher for model
\textit{m10dec} than for our fiducial model \textit{m10}
because the wind particles have not been slowed down by pressure
forces in the inner 
disc. Because of the absence of drag forces within the disc, the
outflow produced by the decoupled wind (Fig.~\ref{fig:m10},
right-hand column) is isotropic
rather than bipolar and the outflow velocity is much higher than for
our fiducial model.

\begin{figure}
\includegraphics[width=0.245\textwidth]{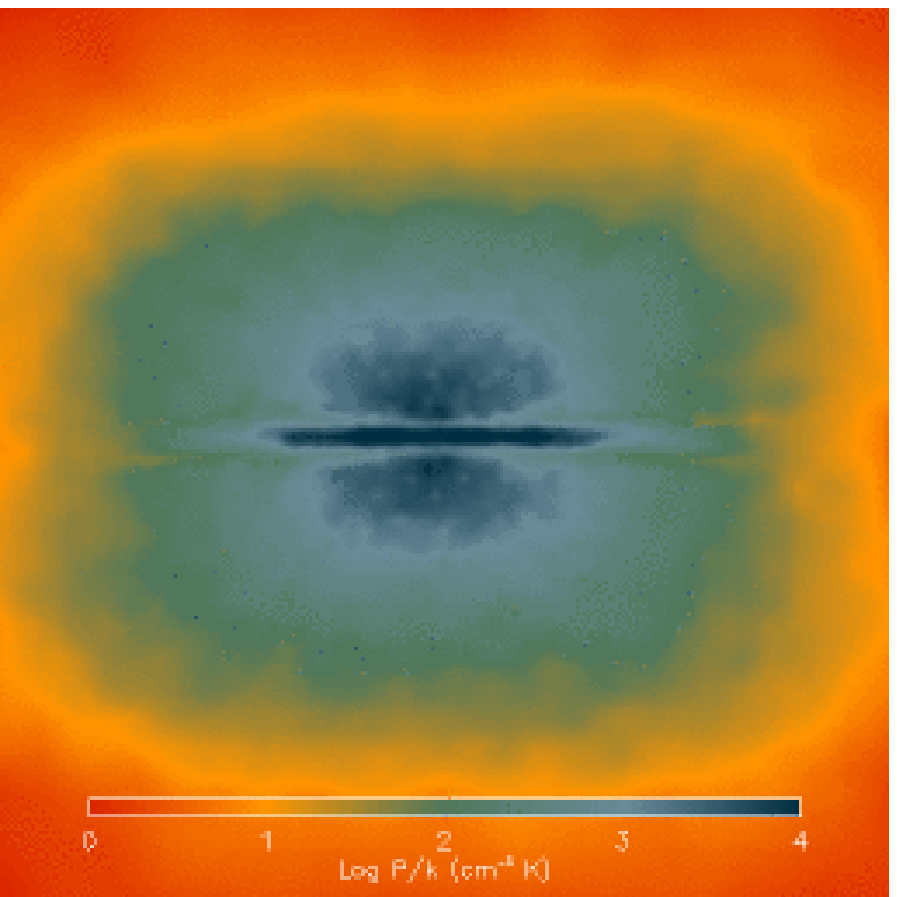}%
\includegraphics[width=0.245\textwidth]{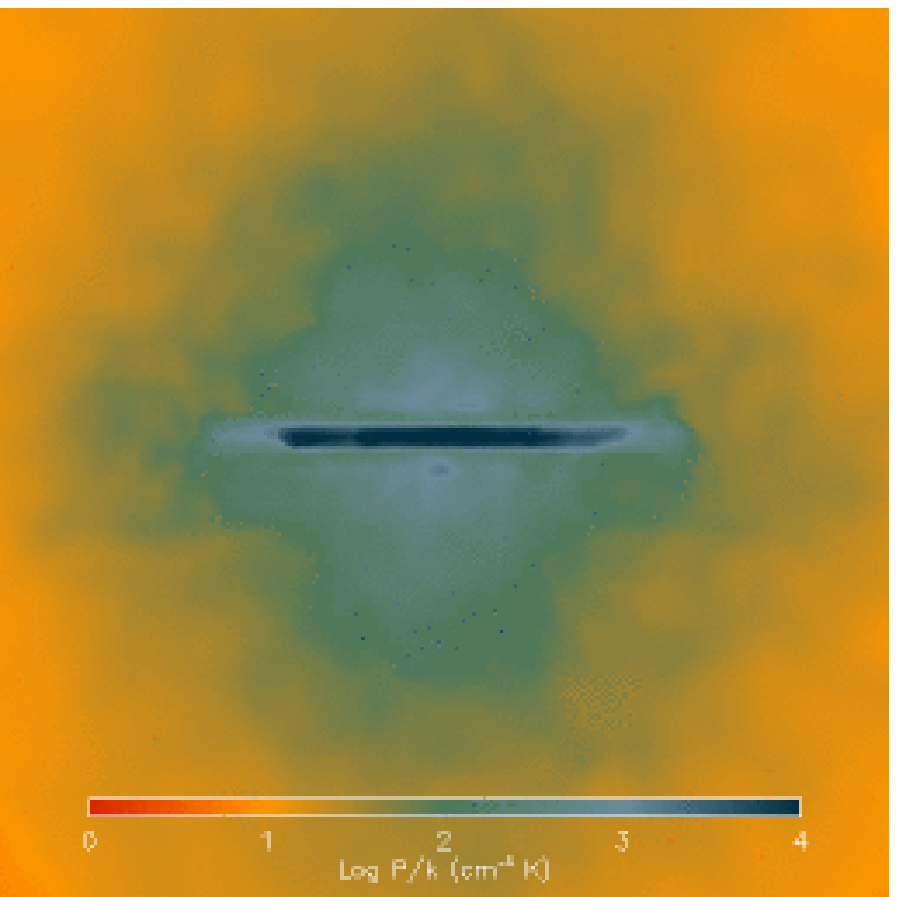}\\
\includegraphics[width=0.245\textwidth]{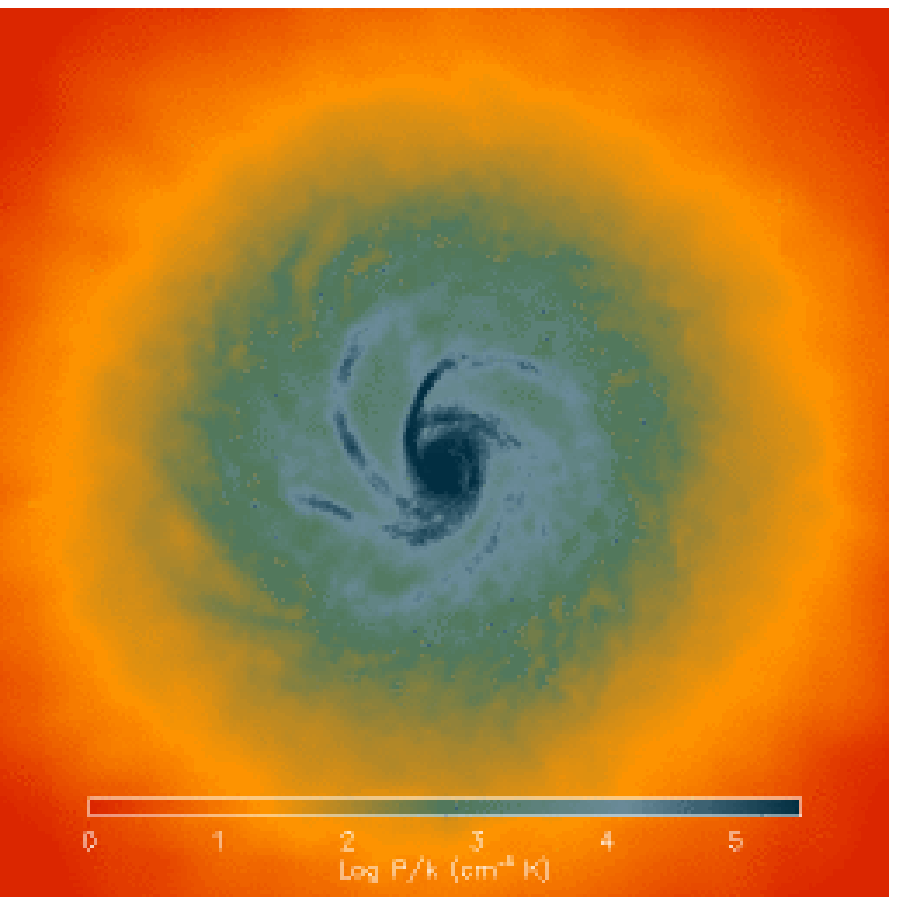}%
\includegraphics[width=0.245\textwidth]{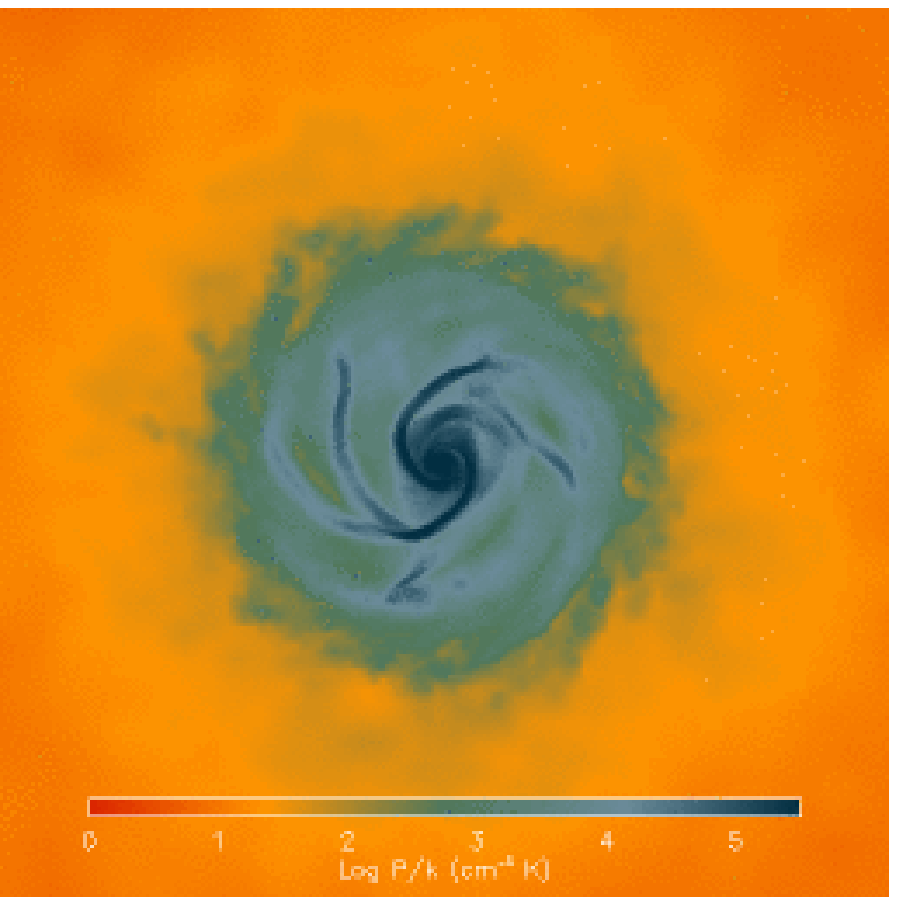}
\caption{Edge- and face-on projections of the gas pressure
  for models \textit{m12} (first column) and \textit{m12dec} (second
  column) at time $t=250~\Myr$. Images $45~\kpch$ on a side. The color coding
  is logarithmic in pressure. The color scale is indicated by the
  color bar in each panel.}
\label{fig:m12pressure}
\end{figure}

As can be seen from the two left-most columns in Fig.~\ref{fig:m12},
the effect of the wind on 
the morphology of our massive galaxy is less dramatic, though still
significant.
The edge-on projection shows that our recipe for
galactic winds results in the creation of a clumpy halo.
As can be seen more clearly from the temperature and pressure plots
(left-hand columns of Figs.~\ref{fig:m12temp} and
\ref{fig:m12pressure}, respectively), the clumps  
are in fact cold, infalling clouds in pressure equilibrium with the
hot wind fluid. The clouds, which are formed through
thermal instabilities, become more
numerous and more prominent during the course of the simulation. 

While the outflow is initially fastest along 
the minor axis, by the end of the simulation the velocities are higher
(though still a factor of a few lower than the initial bi-polar outflow)
for gas flowing out at smaller angles with the
disc. A comparison of Figures~\ref{fig:m12} and \ref{fig:m12pressure}
(left-hand columns) clearly demonstrates 
that the velocity field reflects the gas pressure
distribution. Because the outflow is initially biconical, most of the
gas is deposited (and rains back) along the minor axis, forcing the
gas that is blown out of the disc later to escape along the major
axis. We note that emission profiles will resemble the pressure plots
rather than the velocity field. 

A comparison of
the face-on gas density projections (Fig.~\ref{fig:m12}, bottom row)
indicates that the wind reduces the prominence 
of the spiral arms and suppresses their fragmentation. The wind not only
enhances the stability of the disc, it also creates low-density
bubbles and increases the size of the disc. The increase in the radial
extension of the disc occurs 
partially because gas is pushed outwards by the wind and partially
because some of the gas blown out of the inner disc rains back at
large distances. 

As was the case for the low-mass galaxy, the morphology of the galaxy
that uses hydrodynamically decoupled winds (Fig.~\ref{fig:m12},
right-hand column) looks 
most similar to the model without winds (middle column), except that
the outer disc has been ejected. As we will
show below, the 
reason why the density above and below the disc is lower than in our
fiducial model, is that the gas
is moving through the halo with a much higher velocity. In contrast to
the clumpy galactic fountain produced by our fiducial run, the
decoupled winds result in a smooth and isotropic outflow.

For videos showing the evolution of the gas density, temperature,
and velocity field, we refer the reader to the web site 
\texttt{http://www.strw.leidenuniv.nl/DS08/}. 

\begin{figure*}
\includegraphics[width=0.48\textwidth]{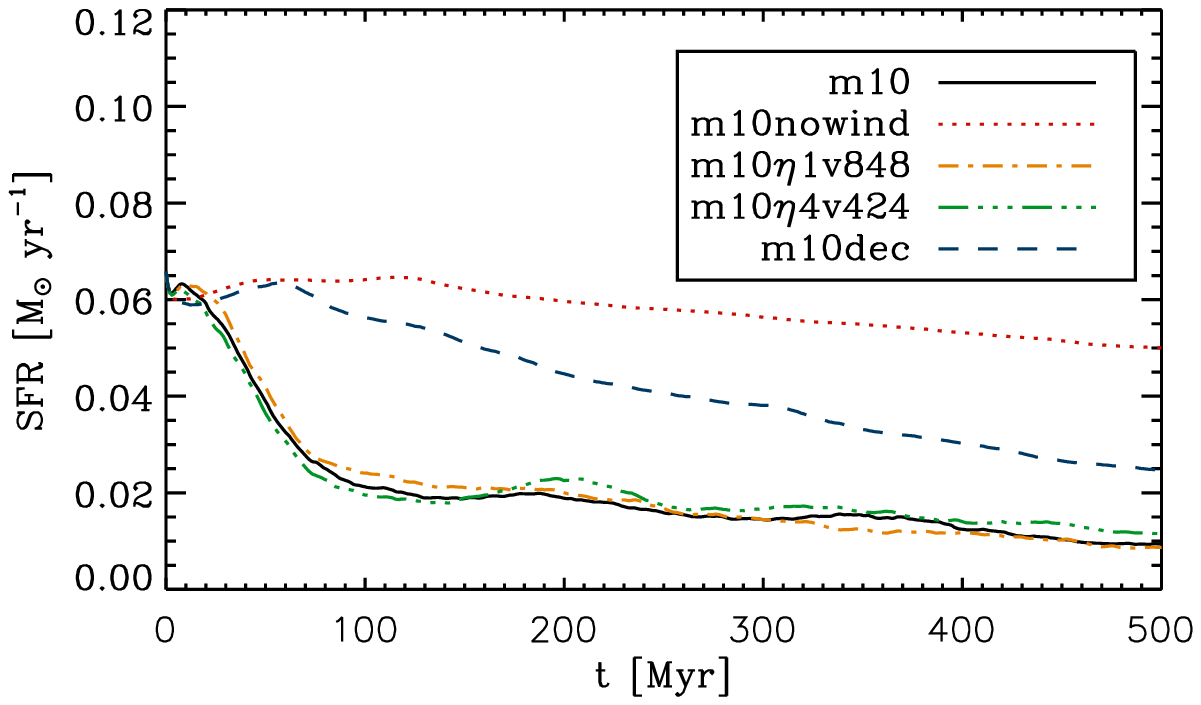}%
\includegraphics[width=0.48\textwidth]{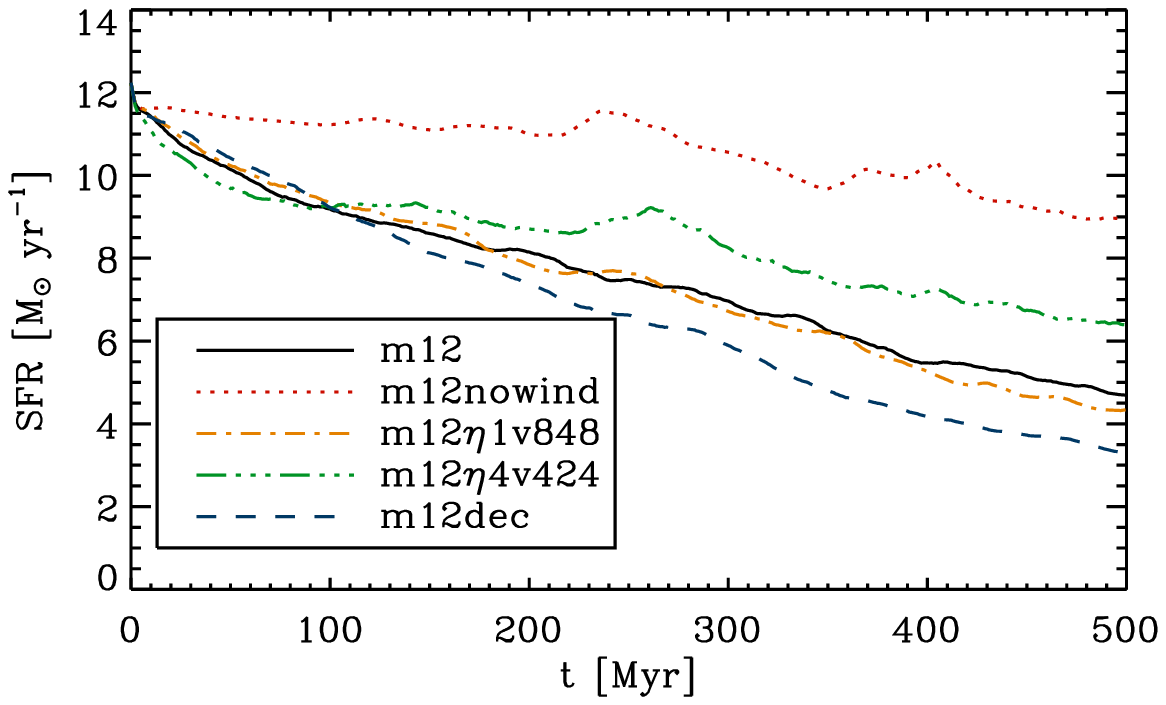}
\caption{Star formation rate as a function of time for the $10^{10}$
  and $10^{12}~\Msolh$ galaxies (left-hand and right-hand panels,
  respectively). Galactic winds strongly reduce the star
  formation rate in all runs that include them. While hydrodynamically
  decoupling the winds reduces their impact on the star formation rate of
  the dwarf galaxy, the opposite is true for the massive
  galaxy. Variations of the mass loading and wind velocity while
  keeping the wind energy per unit stellar mass formed constant have no
  effect except for the lowest velocity in the massive galaxy.
}
\label{fig:sfr}
\end{figure*}

\subsection{Star formation history}

The need to reduce the efficiency of star formation forms the main
motivation for the inclusion of sub-grid recipes for galactic winds in
cosmological simulations. Before demonstrating that our method does
indeed strongly suppress the total star formation
rate (SFR), we note that thanks to our use of the sub-grid model for star
formation presented in \cite{Schaye2008}, all of our models 
agree with the same local, Kennicutt-Schmidt star formation
law.\footnote{The local Kennicutt-Schmidt laws predicted by models
  \textit{m12} and \textit{m12nowind} were shown by \cite{Schaye2008},
  but using a Salpeter IMF rather than the Chabrier IMF used
  here.}

The left-hand panel of Fig.~\ref{fig:sfr} shows the evolution of the
SFR for all high-resolution runs of the low-mass
galaxy. In the absence of galactic winds (dotted curve) the SFR decreases
slowly due to gas consumption. For our fiducial model 
\textit{m10} on the other hand (solid curve), the SFR initially decreases much more 
rapidly before transiting to a more gradual decline after about
100~Myr. For most of the simulation the SFR is about a factor 6 lower
than in model \textit{m10nowind}. Varying the mass loading $\eta$ and
wind velocity $v_{\rm w}$ while keeping the input energy per unit
stellar mass formed ($\propto \eta v_{\rm w}^2$) fixed has no
significant effect on the SFR.

Decoupling the wind particles hydrodynamically (dashed
curve) has a
very large effect on efficiency of the feedback, yielding a SFR intermediate
between those of models \textit{m10} and
\textit{m10nowind}. The fact that the SFR for \textit{m10} is much lower
than that of model \textit{m10dec} may indicate that hydrodynamical drag
increases the effective mass loading to values that are greater than
the input value. 

Interestingly, the high-mass models (right-hand panel) show
qualitatively different behavior. The winds still reduce the star
formation in all runs, but this time the decrease is largest for
the model with hydrodynamically decoupled winds (dashed curve). Apparently,
many of the wind particles that can freely escape in model
\textit{m12dec} are stopped inside the disc due to pressure forces in
the other runs. 

Another difference with the
low-mass galaxy is that the constant wind energy runs no longer give
the same star formation histories. While the two highest wind velocity
runs are still nearly the same, model \textit{m12$\eta$4v424} has a
significantly higher SFR. We conclude that models with a constant wind
energy per unit stellar mass formed yield 
similar star formation histories provided that the wind velocity is
above some minimum value. While this critical velocity
must lie between 424 and $600~\kms$ for the high-mass galaxy, it
must be lower than $424~\kms$ for the low-mass galaxy. The
critical velocity therefore increases with galaxy mass and hence with
the pressure of the ISM.  

Note that when the winds are hydrodynamically decoupled, the SFR will
generally depend on the choice of mass loading even for a fixed wind
energy. In that case the sole effect of the winds is the removal of
fuel for star formation and this happens at a rate that is determined
by the choice of mass loading, provided the wind velocity is
sufficiently high to overcome the gravitational attraction of the
disc.

\subsection{Mass outflow rate and wind velocity}
\label{sec:winds}

\begin{figure*}
\includegraphics[width=0.48\textwidth]{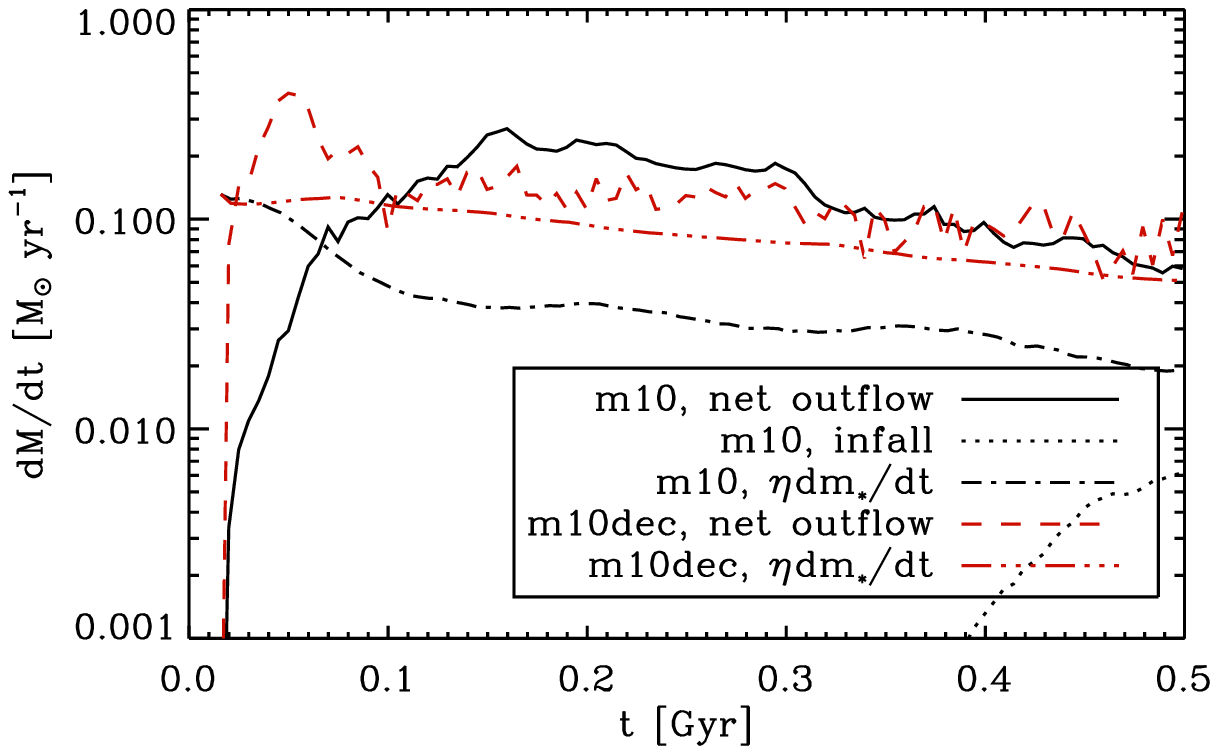}%
\includegraphics[width=0.48\textwidth]{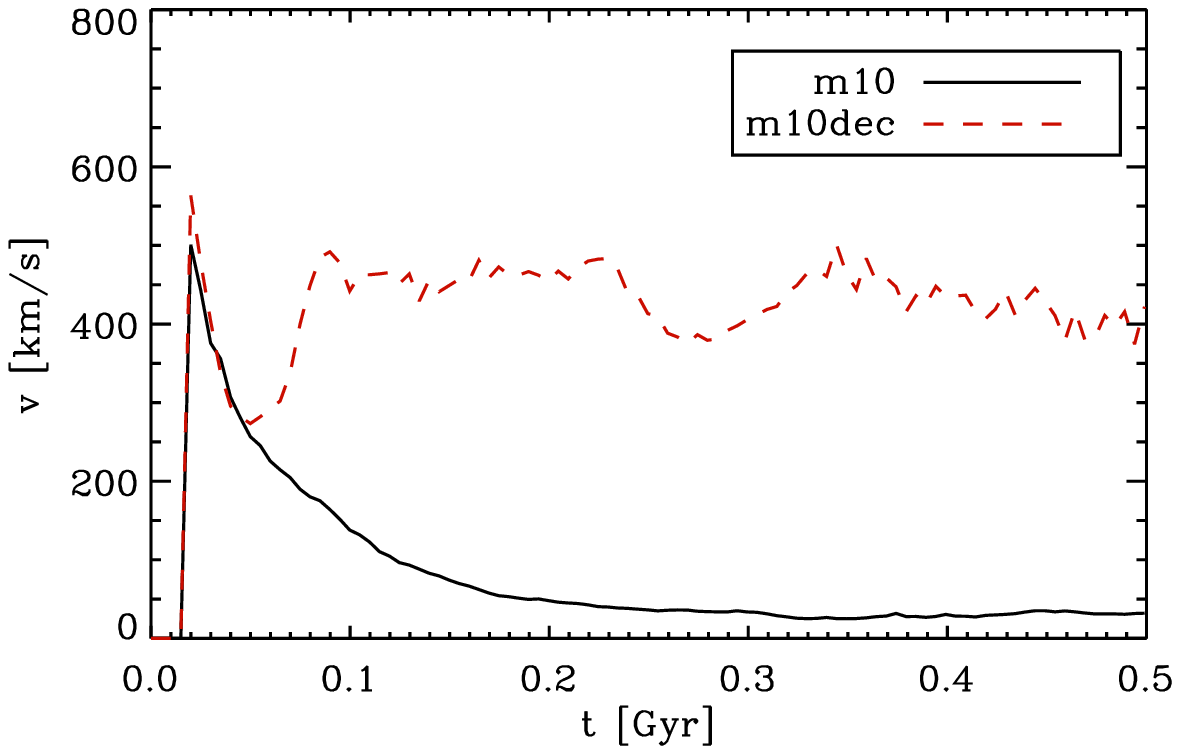}\\
\includegraphics[width=0.48\textwidth]{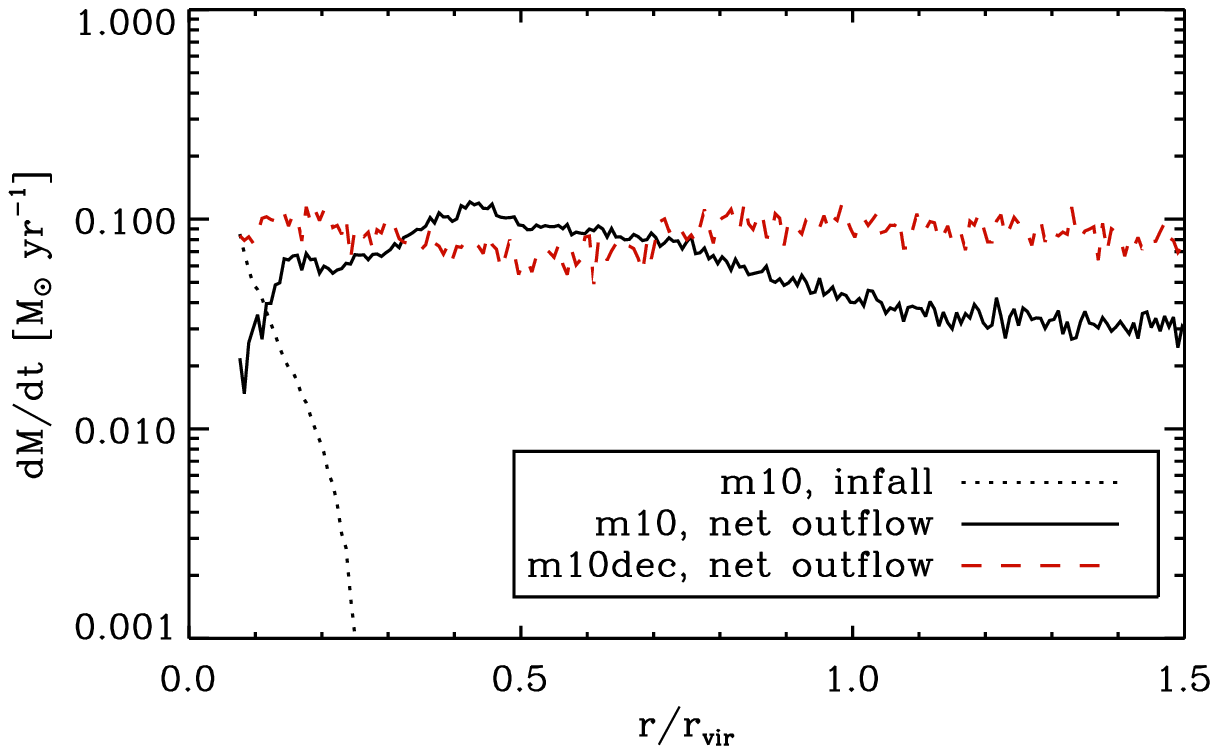}%
\includegraphics[width=0.48\textwidth]{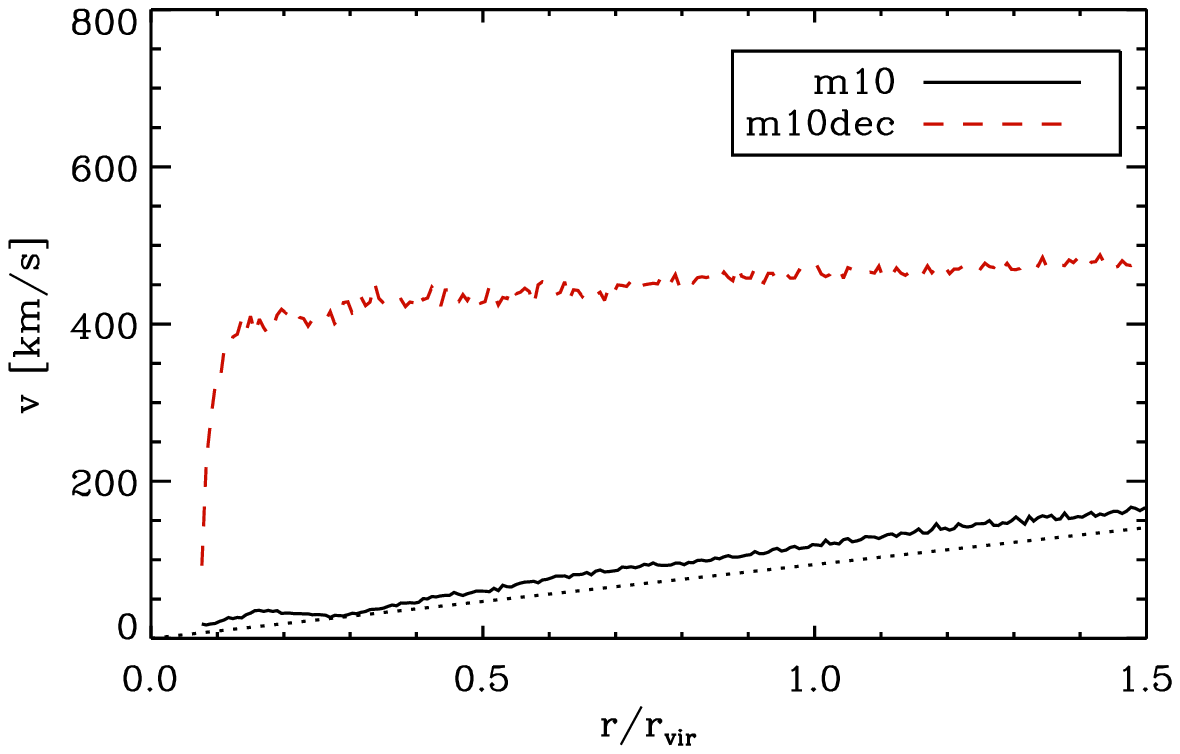}
\caption{Mass outflow rate (left-hand column) and average outflow
  velocity (right-hand column) measured through a spherical shell at
  radius $r=0.2r_{\rm vir}$ as a function of time (top row) and at
  $t=500$~Myr as a function of radius (bottom row) for
  models \textit{m10} and \textit{m10dec}. The dotted line in the
  bottom-right panel indicates the velocity required to reach the
  corresponding radius if the wind were launched from $r=0$ at time
  $t=0$. All other curves are labelled in the legends. Model
  \textit{m10} has a net mass outflow rate that strongly 
  exceeds the input mass loading, whereas the two are similar for model
  \textit{m10dec} in which wind particles temporarily do not feel
  pressure forces. Decoupling the wind particles from the
  hydrodynamics yields much greater outflow velocities.}
\label{fig:m10outflow}
\end{figure*}

Our sub-grid model for generating winds requires two parameters: the
input mass loading $\eta$ and the input wind velocity $v_{\rm w}$. If
the winds are hydrodynamically decoupled, then we expect the
actual wind mass loading and velocity to be similar to the input
values, at least when they are measured at radii small enough for
gravitational deceleration to be unimportant. The situation could be
quite different, however, when pressure forces in the disc are taken
into account. In this section we will investigate this question by
measuring the mass outflow and wind velocity as a function of time and
radius. 

The net mass outflow rate through a surface $S$ is
\begin{equation}
\dot{M} = \int_S \rho \mathbf{v}\cdot{\rm d}\mathbf{S},
\end{equation}
where $\rho$ is the gas density and $\mathbf{v}$ is the gas velocity
relative to the origin, which we take to be the center of the
galaxy. The mass outflow can be split into outflow and 
inflow components according to the sign of the
product $\mathbf{v}\cdot\mathbf{r}$. The net outflow 
rate is then the sum of these two components. We discretize the above
equation as follows 
\begin{equation}
\dot{M}(r,\Delta r)=\frac{1}{\Delta r}\sum_{i=1}^{N_{\rm shell}}m_i
\mathbf{v}_i\cdot\frac{\mathbf{r}_i}{r_i}\,, 
\end{equation}
where $\Delta r$ is the thickness of a spherical shell of radius $r$
centered on the origin (we use $\Delta r=r_{\rm vir}/150$) and $N_{\rm
  shell}$, $m_i$ and 
$\mathbf{r}_i$ are, respectively, the total number, mass and position
of the particles that are within that shell.

The average outflow velocity is
the mass-weighted radial velocity where only particles moving away
from the origin (i.e.\ $\mathbf{v}_i\cdot \mathbf{r}_i > 0$) are included: 
\begin{equation}
\left <v\right >=\frac{\sum_{i=1}^{N_{\rm shell}}m_i(\mathbf{v}_i\cdot\frac{\mathbf{r}_i}{r_i})_+}{\sum_{i=1}^{N_{\rm shell}}m_i},
\end{equation}
where the subscript $+$ is used to indicate that only outgoing particles
are taken into account.

\subsubsection{Models \textit{m10} and \textit{m10dec}}

Fig.~\ref{fig:m10outflow} shows the mass outflow rate (left-hand
column) and the mean outflow velocity (right-hand column) for models
\textit{m10} and \textit{m10dec}. The top row
shows the time evolution 
measured at $r=0.2r_{\rm vir}$, while the bottom row illustrates the
dependence on radius as measured at time $t=500$~Myr. 

Let us start with the evolution of the mass flux (top-left panel). 
After a rapid, initial rise, the net mass outflow rate in model
\textit{m10} (solid curve) decreases smoothly with
time. By the end of the simulation infall starts to become significant
(dotted curve). Note that since there was no gaseous halo in the
initial conditions, all of the infalling gas must have been blown out
at earlier times. It is interesting to compare the net outflow rate
with the input mass outflow rate $\eta \dot{M}_\ast(t')$ (dot-dashed
curve) which we evaluate at the retarded time $t' = t-r/v_{\rm
  w}$. Clearly the net mass flux is much greater than the input
value. This confirms that most wind particles drag many other gas
particles along. 

The dashed and dot-dot-dot-dashed curves correspond to the actual and
input net mass loading for model \textit{m10dec}. As expected, they
agree much better than for model \textit{m10}. When pressure forces
are temporarily turned off, the wind particles can freely escape into
the halo and there is no opportunity to change the wind mass
loading. Even for this model the net mass loading slightly exceeds the
input value, however, which implies that at the time when pressure forces are
turned back on (i.e.\ when the density falls below one tenth of the
star formation threshold) at least some wind particles are still in
the outer layers of the disc and can therefore drag some other
particles along. This also explains why the disc in model
\textit{m10dec} is somewhat smaller and thinner than in model
\textit{m10nowind} (see Fig.~\ref{fig:m10}). The net mass outflow rate
is, however, much smaller than for model \textit{m10}, which explains
why the hydrodynamically decoupled winds yield a much smaller
reduction of the SFR than our fiducial wind model (see
Fig.~\ref{fig:sfr}). 

The top-right panel of Fig.~\ref{fig:m10outflow} shows the evolution
of the mean outflow velocity. After a sudden rise, the
velocity drops to a steady value of slightly less than $50~\kms$, far
below the input value of $600~\kms$. We note, however, that there is
significant scatter around this value and that some particles have
much higher velocities, particularly around the minor axis (see
Fig.~\ref{fig:m10}). The mean outflow velocity is low because a
typical wind particle has to penetrate the large amount of gas blown
out at the beginning of the simulation before it can
escape into the vacuum (see Fig.~\ref{fig:m10}). 

The mean outflow velocity in run \textit{m10dec} is about $400~\kms$,
which exceeds that of \textit{m10} by an order magnitude. The outflow
velocity is smaller than the input value $v_{\rm w} = 600~\kms$
because of projection effects and because some hydro drag is 
felt in the outer layers of the disc, not because of gravitational
deceleration. 

\begin{figure*}
\includegraphics[width=0.48\textwidth]{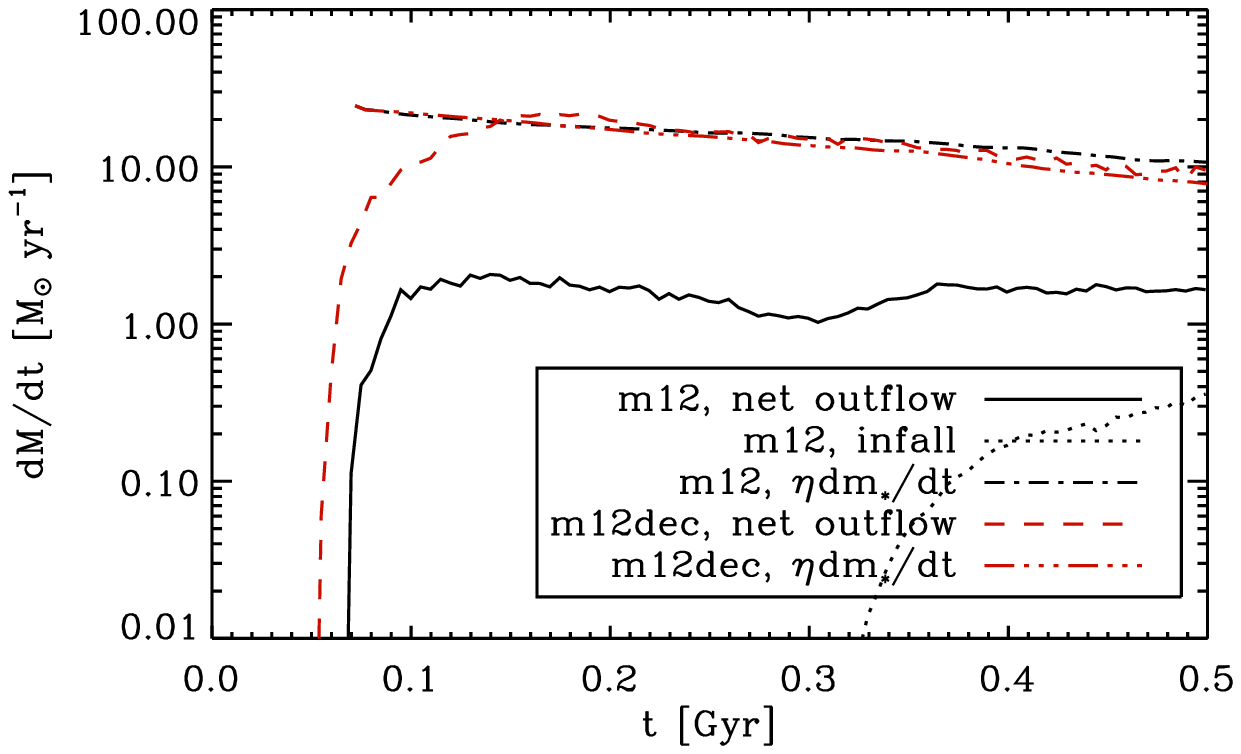}%
\includegraphics[width=0.48\textwidth]{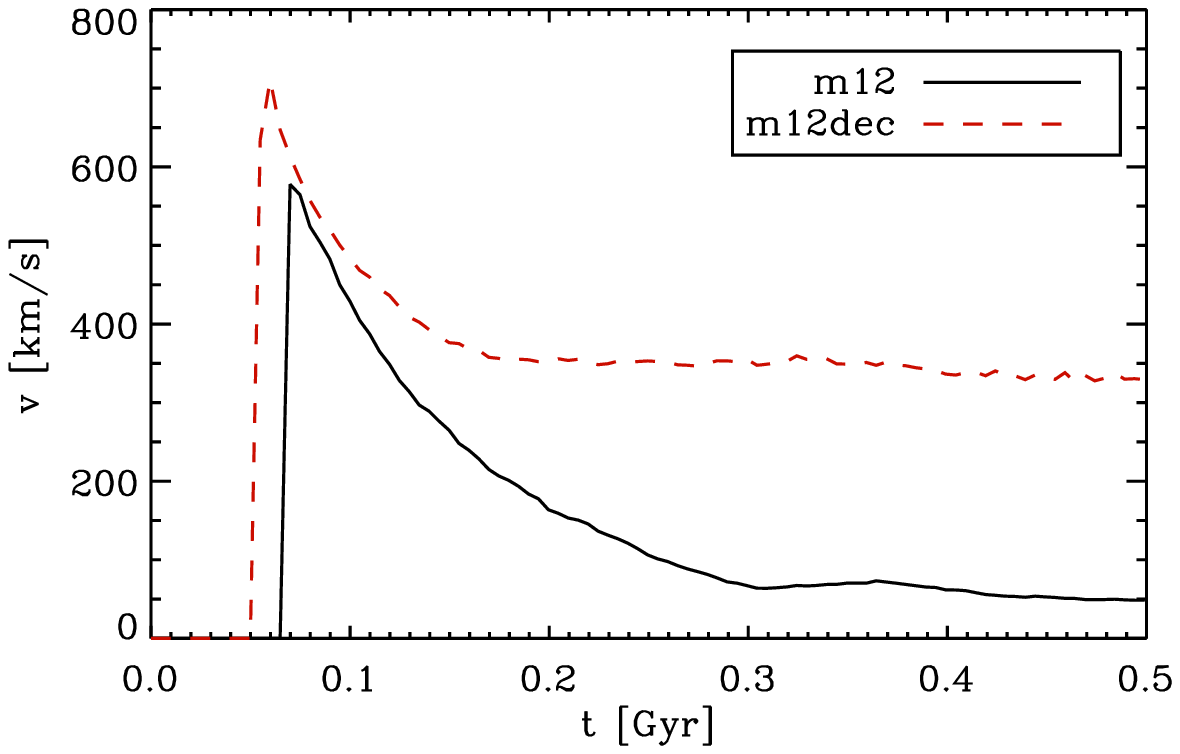}\\
\includegraphics[width=0.48\textwidth]{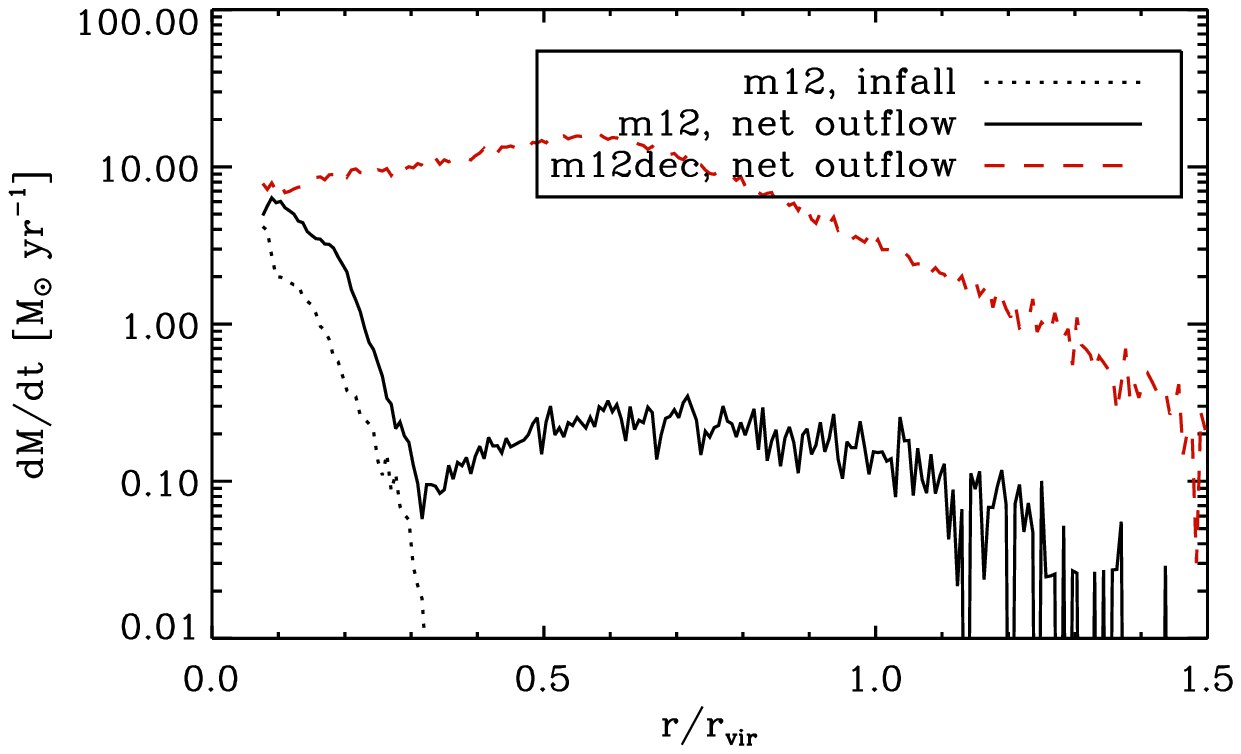}%
\includegraphics[width=0.48\textwidth]{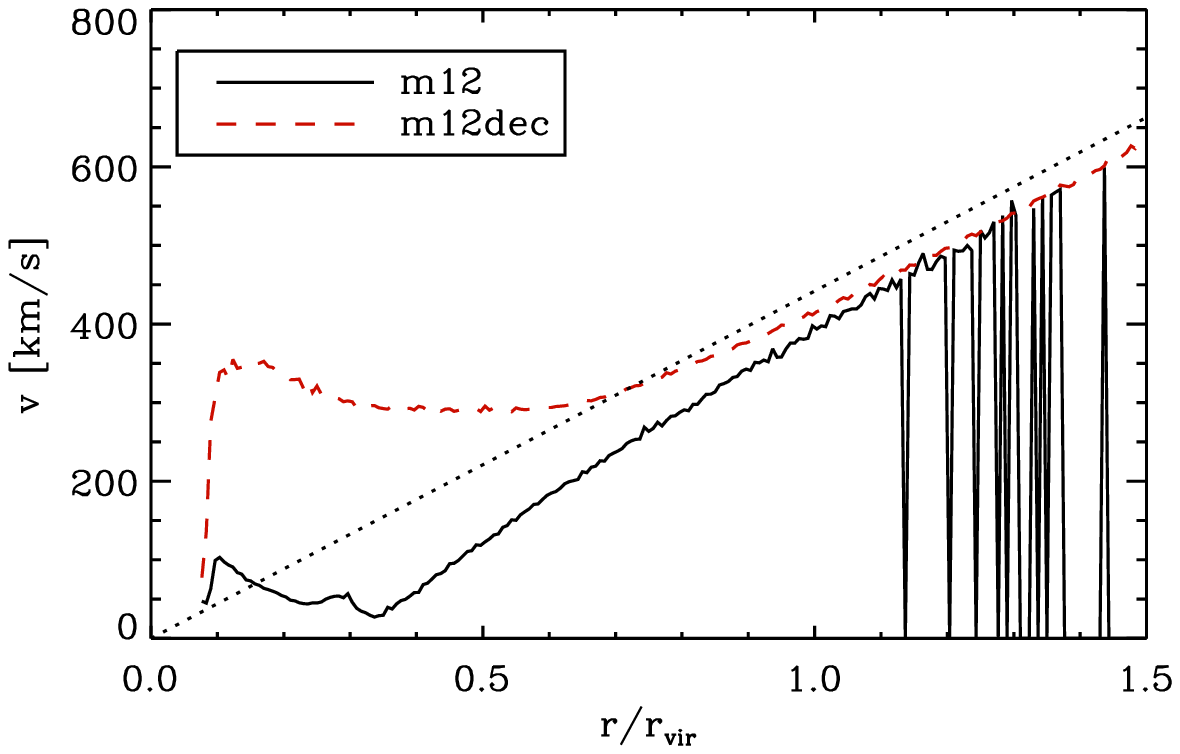}
\caption{As Fig.~\protect\ref{fig:m10outflow} but for models
  \textit{m12} and \textit{m12dec}. Contrary to the result for
  \textit{m10}, the net mass loading for model \textit{m12} is much
  lower than the input 
  value. When the winds are hydrodynamically decoupled, the actual
  value again agrees with the input value. Decoupling the wind
  particles from the hydrodynamics yields much greater outflow
  velocities. The radial dependence of the outflow velocities tracks
  the dotted line, which implies that the increase with radius is due
  to the finite time that has elapsed since the launch of the wind.}
\label{fig:m12outflow}
\end{figure*}

The outflow velocity in \textit{m10} is in perfect agreement
with observations of dwarf galaxies with similar circular velocities
and star formation rates \citep{Schwartz&Martin2004}. In contrast,
outflow velocities as large as 
predicted by model \textit{m10dec} have only been observed in galaxies
that have star formation rates exceeding that of \textit{m10dec} by
two orders of magnitude
\cite[e.g.][]{Rupke2005,Martin2005,Veilleux2005}. This suggests 
that pressure forces within the 
disc are  instrumental in shaping the winds emanating from dwarf
galaxies. 

The bottom row of Fig.~\ref{fig:m10outflow} shows how the mass
outflow rate and wind velocity vary with radius at time $t=500$~Myr,
which marks the end of the simulations. The net mass outflow
rates (bottom-left panel) are nearly independent of radius. For model
\textit{m10} infall 
becomes significant at $r < 0.2 r_{\rm vir}$ (dotted curve) but it is
negligible at all radii for \textit{m10dec} (not shown). 

While the
outflow velocity is nearly constant for model \textit{m10dec} (bottom-right
panel, dashed curve), it shows a strong increase with radius for
model \textit{m10} (solid curve). The dotted line indicates the
velocity with which 
a particle needs to travel to reach the corresponding radius after
500~Myr. The actual outflow velocity can exceed this value if the gas
was launched less than 500~Myr ago or it can fall
below the dotted line if the wind 
has decelerated and/or if it was launched from radii greater than
zero. Model \textit{m10} roughly tracks the dotted line, which implies
that the increase in the outflow velocity with radius is a trivial
consequence of the 
combination of a relatively small outflow velocity and a finite time
interval since 
the launch of the wind. In other words, for each radius the wind is
close to the minimum value it must have in order to reach that
radius. Note that a similar trend may well be present in real starburst
galaxies, provided that they are observed within a few hundred million
years after a sharp rise in their star formation rates.

\subsubsection{Models \textit{m12} and \textit{m12dec}}

Fig.~\ref{fig:m12outflow} is identical to
Fig.~\ref{fig:m10outflow} except that it shows the results for the
$10^{12}~\Msolh$ galaxy. While many of the results are qualitatively
similar to those for the $10^{10}~\Msolh$ galaxy, there are also some
noticeable differences. The most important difference is that the net
mass outflow for model \textit{m12} is an order of magnitude below the
input value (top-left panel, solid and dot-dashed curves), whereas it
exceeded the input value in the case of \textit{m10} (see
Fig.~\ref{fig:m10outflow}). Apparently, most 
of the wind particles are halted by pressure forces before they can
escape the disc of the massive galaxy. 

In contrast, when pressure forces are temporarily switched off, the
actual and input mass loading agree very well (dashed and
dot-dot-dot-dashed curves).  This explains why, for the massive
galaxy, the decoupled winds
are more effective at reducing the SFR than the winds
in our default model and why the situation is reversed for the
dwarf galaxy (see Fig.~\ref{fig:sfr}). The pressure of the ISM in
dwarf galaxies is too low to quench the wind, allowing it to drag
large amounts of gas out of the disc. In high-mass galaxies the ISM
pressure must be greater in order to support the larger surface
densities and the main effect of hydrodynamical drag is to stop the
wind from escaping the disc. 

We note, however, that the ability to completely quench the wind,
i.e., to stop the wind particles before they have moved significantly,
must depend on the ram pressure exerted by individual wind 
particles and hence on the input wind
velocity. Thus, models with identical input wind energies, but higher
wind velocities may produce larger mass outflows for massive
galaxies. Fig.~\ref{fig:sfr} indicates that increasing the wind
velocity from 600 to $848~\kms$ for our $10^{12}~\Msolh$ galaxy does
not make a difference, whereas decreasing it to $424~\kms$ reduces the
impact of the wind. This could be largely a numerical effect since the ram
pressure exerted by individual wind particles depends on their
mass. Fig.~\ref{fig:sfr} demonstrates that once the input velocity
is above some critical value, which increases with galaxy mass,
the results become independent of the way in which the available wind
energy is distributed. 

The bottom-right panel of Fig.~\ref{fig:m12outflow} illustrates other
differences between the high- and low-mass
galaxies. At a fixed fraction of the virial radius, the outflow velocities
are higher than for the dwarf galaxy. Furthermore, for $r\ga r_{\rm vir}$
not only \textit{m12} but also \textit{m12dec} track the dotted line
which indicates the radius that a wind particle launched with the
corresponding velocity can reach within 500~Myr. All these differences
can be explained by the fact that a fixed
fraction of the virial radius corresponds to a greater physical radius
for a more massive galaxy and that a higher velocity is required to
reach a greater physical radius in a 
fixed amount of time.

\subsection{Resolution tests}
\label{sec:resol}

We performed resolution tests for our fiducial runs, decreasing the
particle numbers by factors of 8, 64 and 512. Hence, the particle
mass is increased by the same factors whereas the SPH smoothing kernels
are increased by factors 2, 4 and 8,
respectively (for a fixed density). Note that the lowest-resolution
runs have only 459 particles in the disc and therefore do not
provide enough samples to measure the wind properties. 

As discussed in 
section~\ref{sec:code}, for star-forming gas in
model \textit{m12} the ratio of the kernel mass to the Jeans mass is
$1/6$ and the ratio of the kernel size to the Jeans length is
$1/(48)^{1/3}$ (recall that these ratios are constant because
star-forming particles follow a power-law effective equation of state
with polytropic index $\gamma_{\rm eff}=4/3$). Hence, the Jeans scales are
not resolved by any of the  lower
resolution versions of the $10^{12}~\Msolh$ galaxy, whereas they are
resolved by all but the \textit{m10lr512} version of the
$10^{10}~\Msolh$ galaxy. 

\begin{figure*}
\includegraphics[width=0.48\textwidth]{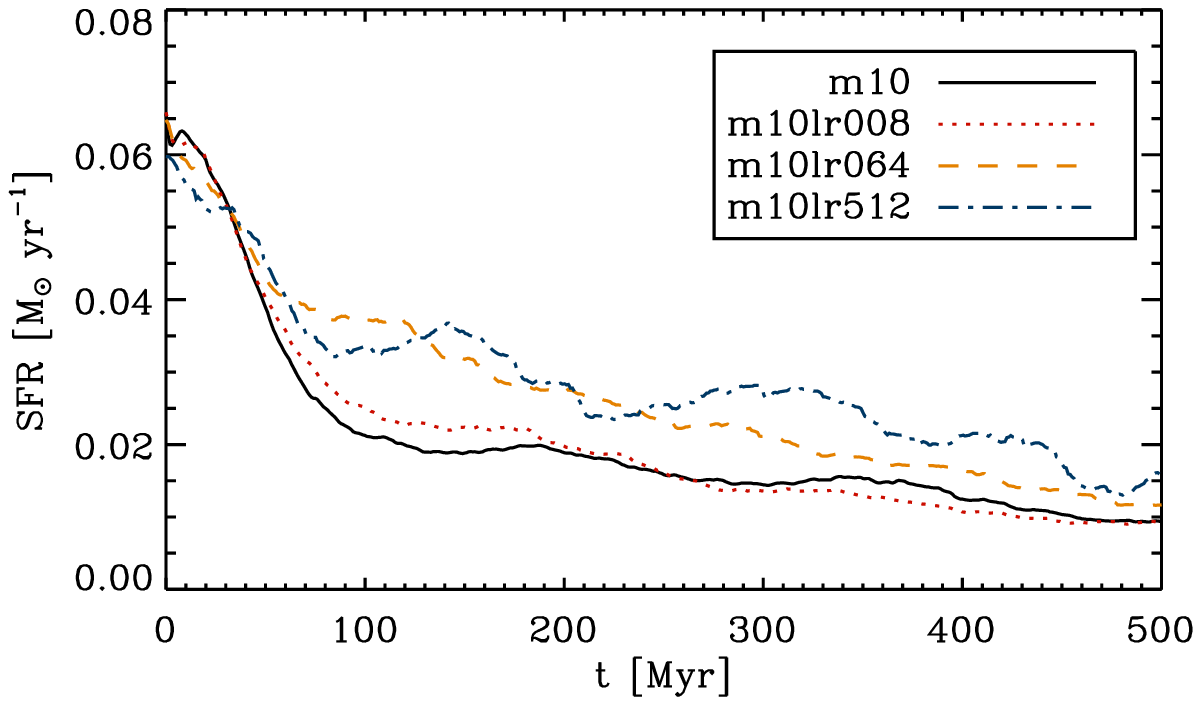}
\includegraphics[width=0.48\textwidth]{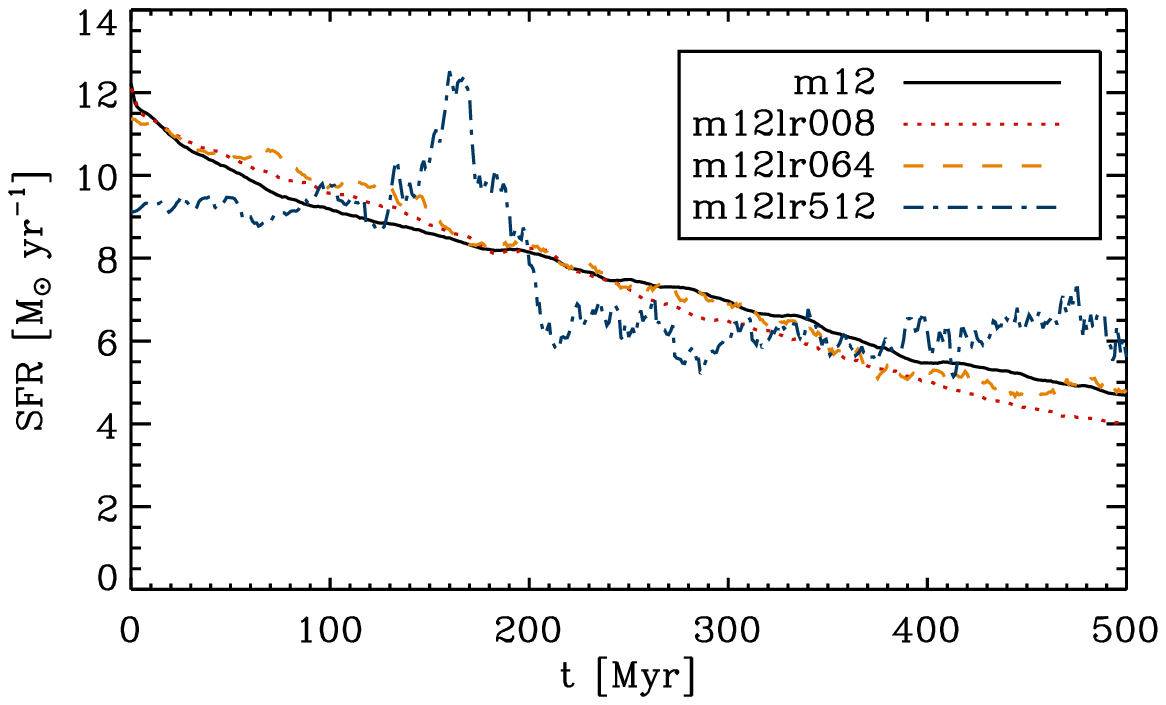}
\caption{Resolution dependence of the star formation rate as a
  function of time for the $10^{10}$ and $10^{12}~\Msolh$ galaxies
  (left-hand and right-hand panels, respectively). Resolution has only
 modest effects, except for model \textit{m12lr512} which becomes
 violently unstable.}
\label{fig:sfr_res}
\end{figure*}

\begin{figure*}
\includegraphics[width=0.48\textwidth]{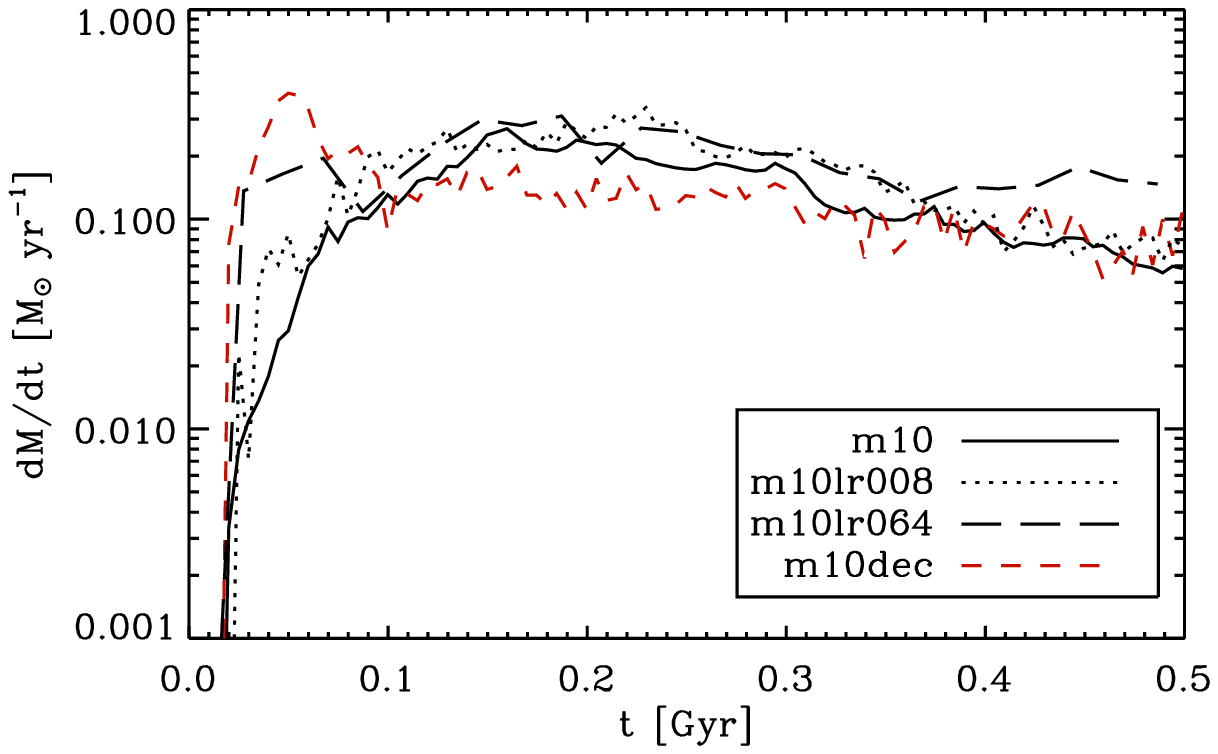}%
\includegraphics[width=0.48\textwidth]{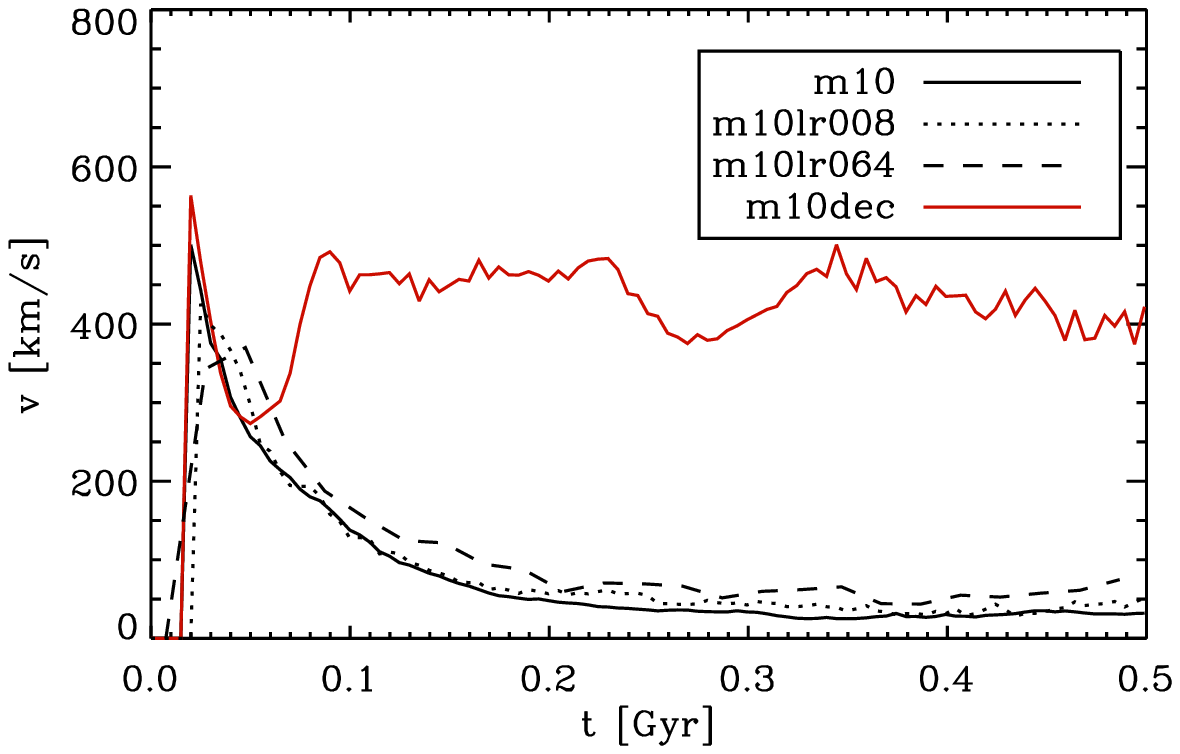}\\
\includegraphics[width=0.48\textwidth]{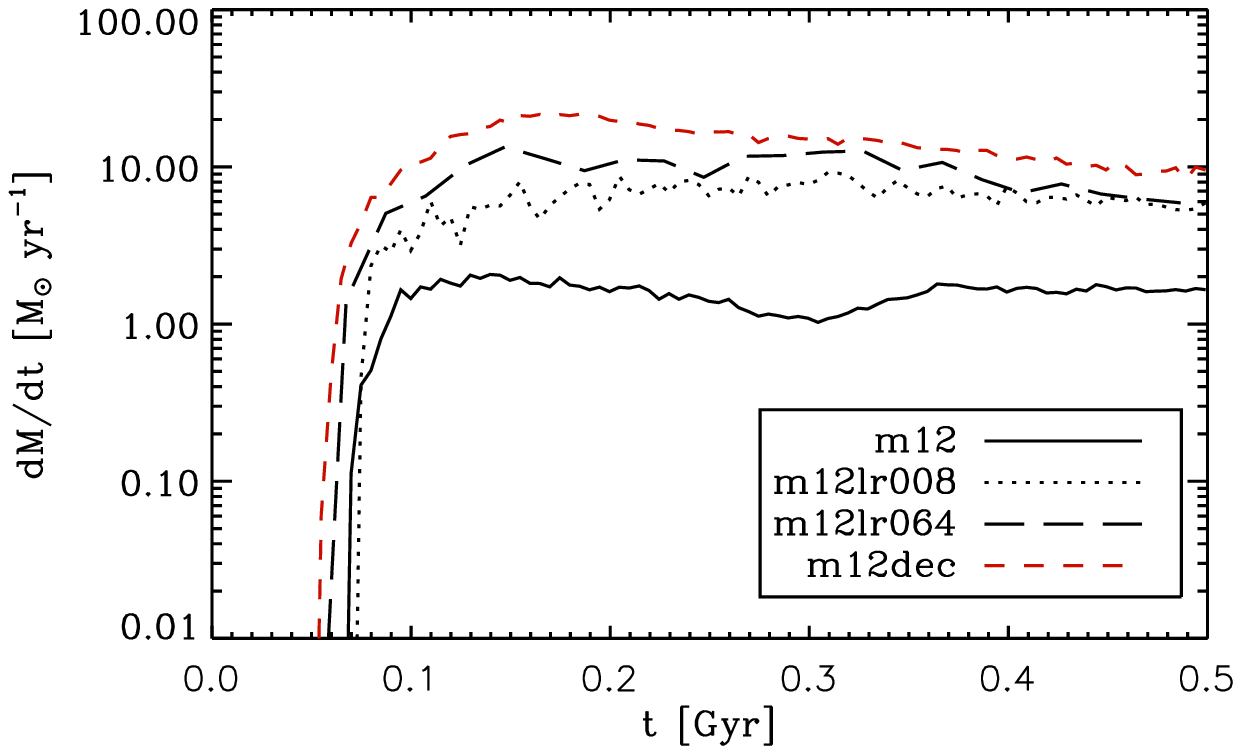}%
\includegraphics[width=0.48\textwidth]{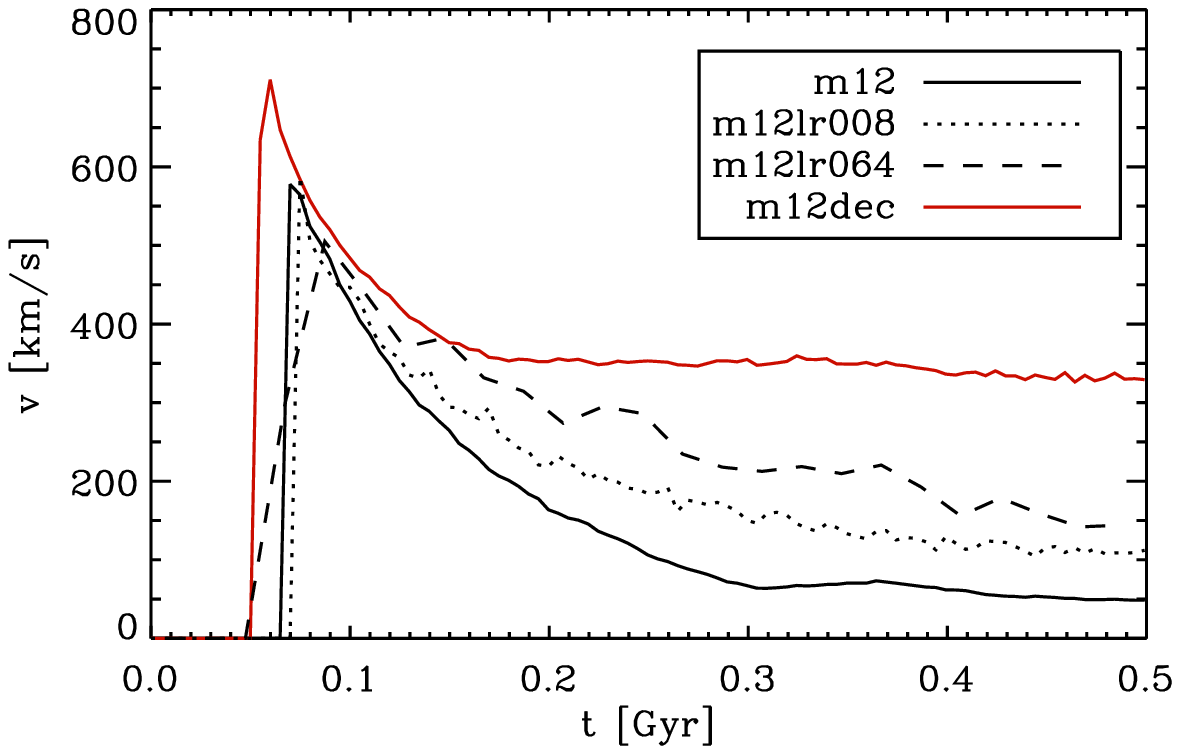}
\caption{Resolution dependence of the mass outflow rate (left-hand
  column) and mean outflow velocity (right-hand column) measured
  through a spherical shell at radius $r=0.2r_{\rm vir}$ as a
  function of time for the $10^{10}$ (top row) and the
  $10^{12}~\Msolh$ (bottom row) galaxies. Decreasing the resolution
  yields higher mass outflow rates and higher outflow velocities for the
  high-mass galaxy, but has no discernible effect on the low-mass
  galaxy. Note that the particle mass in \textit{m10lr064}, which is
  the lowest-resolution version plotted, is still lower than that in
  \textit{m12}.}
\label{fig:resolution}
\end{figure*}

Fig.~\ref{fig:resolution} confirms that
convergence requires resolving the Jeans scales. Interestingly,
Fig.~\ref{fig:sfr_res} demonstrates that high resolution is much
less critical for obtaining a resolved SFR than it is
for obtaining a converged estimate of the wind properties. This is
probably due to the fact that the SFR is determined by
the gas distribution in the disc, which is initially identical for all
runs, whereas the wind behaviour is rather sensitive to
  how well the vertical scale of the disc is resolved.
The
situation would likely be quite different for an ab initio run using
cosmological initial conditions. In such a simulation a galaxy
consisting of only a few hundred particles would probably not even
form a disc. 

As the resolution is decreased, the net mass outflow rate and the wind
velocity both increase for the high-mass galaxy. As expected, they
remain the same for the low-mass galaxy. Thus, unresolved simulations
give results 
that are more similar to the models in which the wind particles are
temporarily decoupled hydrodynamically. However, even for the lowest
resolution models for which we can still make meaningful measurements,
the wind mass loading and outflow velocity are below those of model
\textit{m12dec}.\footnote{If the
  wind particles are decoupled hydrodynamically, the results are
  nearly independent of resolution.}

The fact that decreasing the resolution mimics the effect of
decoupling the hydrodynamics is not surprising. In the limit of only
one particle per scale height, most wind particles will not encounter
any other disc particles on their way out, leaving only the drag
exerted by their original SPH neighbours.

\section{Summary and discussion}
\label{sec:disc}

Feedback from star formation is thought to play a key role in
determining the observed properties of galaxies. It is therefore
essential to include it in simulations of 
galaxy formation and evolution. Current simulations lack the
resolution to resolve the energy conserving phase in the evolution of
supernova remnants, causing any thermal energy input to be mostly
radiated away before it has any significant hydrodynamical effect.
Several types of sub-grid
prescriptions for the generation of galactic winds have
been proposed with the intent of overcoming numerical limitations. The
most widely used methods use either thermal
feedback combined with a temporary suppression of radiative
cooling or kinetic feedback.

In this work we introduced a sub-grid recipe to model feedback from
massive stars in cosmological SPH simulations. The energy is
distributed in kinetic form among the neighbours of recently formed
stars. We implemented our prescription in the SPH code \textsc{gadget} and
tested it using high-resolution simulations of isolated disc galaxies
of total mass $10^{10}$ and $10^{12}~\Msolh$. Our kinetic feedback scheme
strongly reduces the star formation rates and has a dramatic
impact on the morphology of the galaxies. 

The disc of the dwarf galaxy
becomes puffed up and punctuated with low-density bubbles, resembling the
HI observations of nearby galaxies. A bipolar outflow develops
naturally, with the largest velocities along the minor axis. The end  
result is a rather diffuse and irregular galaxy. Winds make the  
disc of the massive galaxy more stable, yielding more diffuse
and less fragmented spiral arms. Its outer parts become more extended
and also contain a large number of bubbles. While the wind is
initially fastest along the minor axis, after a few hundred million 
years the velocities are higher at large opening angles because of
strong infall along the minor axis. The edge-on gas density,
temperature and pressure profiles remain, however, highly
bi-conical. The infall is highly clumpy, consisting of cold gas clouds
that have formed through thermal instabilities in the hot wind.

The mean outflow velocity decreases with time, because gas that 
is blown out of the disc later must plough through gas
that was blown out earlier, some of which has
turned around and is falling back. At large radii the velocity
increases with radius due to the fact that only gas with some minimum
velocity can reach a given radius in the time since the initial
starburst. The mean
outflow velocities are much lower than the input 
values ($600~\kms$ for our fiducial model), except at the largest
radii reached by the wind.  

While the net mass outflow rate exceeds the input value ($
\dot{M}_{\rm w} / \dot{M}_\ast=2$ for our fiducial model)
by about an order of magnitude for the dwarf galaxy, the reverse is
true for the massive galaxy. Apparently, (ram) pressure forces in the
disc enable the wind particles to drag large amounts of ISM out of the
dwarf galaxy's disc, but are able to confine most wind particles to
regions close to the disc in the case of the massive galaxy, whose ISM
has both a higher density and pressure. 

We varied the input wind velocity from 424 to $848~\kms$ while
keeping the total kinetic energy per unit stellar mass formed constant
by adjusting the input mass 
loading accordingly. For the dwarf galaxy the resulting star formation histories
are all nearly identical and in the case of the massive galaxy only
the lowest velocity run differed significantly. Thus, provided
the input wind velocity is 
higher than some minimum value, which increases with the mass (and
thus ISM pressure) of the galaxy, the results depend on the input kinetic
energy but are insensitive to the amount of mass the energy is
distributed over. 

While the star formation histories are relatively insensitive to
numerical resolution, convergence of the predictions for the outflows
requires resolving the Jeans scale.

We contrasted our scheme with the sub-grid model 
of SH03 which has been widely used in the literature. In the SH03
prescription, the wind particles are selected stochastically from all
the star-forming (i.e.\ dense) particles in the simulation and are
therefore non-local. These wind particles are subsequently decoupled
from the hydrodynamics for 50~Myr (i.e.\ 31~kpc if traveling at
$600~\kms$) or until their density has fallen
below 10 percent of the threshold for star formation.

In this work
we have not tested the effects of non-locality because we expect them
to be small for high-resolution simulations of individual
galaxies such as those presented here. However, we do expect significant
differences in cosmological simulations for which most galaxies will contain
only a small number of star particles. In such galaxies the
formation of star particles and the injection of kinetic energy will
become essentially uncorrelated locally. To see this, note that for a
galaxy that contains only a few star-forming particles, the typical
time difference between the 
kicking of a wind particle and the creation of a star particle will be of 
order the gas consumption time scale divided by the input mass
loading, or $\sim 10^9/\eta~\yr$ at the  
threshold for star formation. This time scale greatly exceeds
both the lifetime of massive stars and the simulation time step. The
disconnection between the creation of a star particle and the
injection of kinetic energy in its surroundings may have some
undesirable consequences. For example, predictions for the chemical
enrichment of the 
intergalactic medium may be affected because wind particles
will contain an amount of heavy elements typical of particles in their
host galaxies (which may often be zero) rather than abundances typical
of the gas surrounding newly-formed stars. 

We carried out simulations in which the wind particles were
temporarily decoupled from the hydrodynamics as in the SH03 prescription. The
difference between the predictions for the fiducial and decoupled
models is dramatic. Decoupled winds have almost no effect on the
morphology of the disc. Compared with the fiducial model and
HI observations of nearby galaxies, the dwarf
galaxy, which has a low enough surface density to be stable, is much
smoother while the massive galaxy, which is unstable without the
injection of turbulence, is more clumpy but lacks low-density
bubbles. While the winds in the fiducial model slightly increase the
size of the gas disc, the decoupled winds continuously shrink the
disc. While the coupled winds drive a
large-scale bipolar outflow from 
the dwarf galaxy and a clumpy galactic fountain in the
massive galaxy, the decoupled winds produce isotropic outflows in both
cases. 

For the decoupled winds the 
outflow velocities (at least for $r < r_{\rm vir}$) are constant in
time and nearly independent of the galaxy mass. The mean, projected outflow
velocity is about 70~percent of the input value, which greatly exceeds
the outflow velocities observed in starbursting dwarf galaxies. The
net mass outflow rates are in good agreement with the input
values. The kinetic energy in the wind escaping the disc is orders of
magnitude higher than in the case of the coupled winds. 
Compared with the fiducial model, the decoupled winds are less
efficient at suppressing the SFR in the dwarf galaxy,
but more efficient for the massive galaxy. The wind properties in the
decoupled runs are insensitive
to numerical resolution, even when the Jeans scale is completely
unresolved. The lower the resolution of the runs with coupled winds,
the more they resemble the case of decoupled winds.

SH03 had several motivations for decoupling the winds
hydrodynamically. They wanted to calibrate the sub-grid model with
observations outside the disc, because they had no hope of resolving
the structure of the ISM in their cosmological simulations, and they
wanted a recipe that was insensitive to numerical resolution. Our
tests clearly demonstrate that their method satisfies both these
requirements. 

One may question, however, whether the observational uncertainties are
not far too large to enable a 
calibration of the wind velocity and mass loading outside the disc. It is
currently not even clear how the observed values depend on radius and
gas phase. It is also questionable whether it is desirable
for the predictions of hydrodynamical simulations to be insensitive to
resolution if the simulations do not resolve the Jeans scale.

Even if one is not
interested in the internal structure of galaxies, which agrees less well with
observations if the winds are decoupled, there are likely to be
important differences in other types of predictions. The fact that
hydrodynamic drag makes low-mass galaxies much more diffuse may, for
example, greatly affect predictions for quasar absorption line
observations and may also alleviate the angular momentum problem
(i.e.\ simulated 
discs are too small, probably due to excessive transfer of angular
momentum to the dark matter halos). The fact that neglecting pressure
forces on wind particles within the disc increases the kinetic energy
of the escaping gas by very large factors, 
may have a large impact on predictions for the chemical
enrichment of the intergalactic medium.

However, we stress that our
simulations lack the resolution and the physics
necessary to predict the 
structure of the multiphase ISM and to model the small-scale effects
that ultimately lead to the development of galactic winds. It is also
important to note that while our 
artificial set-up of isolated, thin discs is useful for numerical
experiments such as those presented here, it may exaggerate the
differences between the coupled and decoupled winds. Galaxies in
cosmological simulations are surrounded by gaseous halos and in the
SH03 prescription wind
particles will be re-coupled to the hydrodynamics before they
leave the halos. 

Summarizing, our results suggest that (ram) pressure forces in the
disc and the inner halo have a very strong impact on the structure of the
ISM and the properties of galactic winds. Pressure forces exerted by
expanding superbubbles puff 
up disc galaxies, give low-mass starbursting galaxies 
irregular morphologies and stabilize the discs of massive
galaxies. The energy lost in this process strongly reduces the
kinetic energy carried by the outflowing gas. Even if the first bubbles open
up a channel in the disc through which the gas can be efficiently
ejected, it will run into the gas which was blown out earlier and has
been decelerated in the process. For massive galaxies
the reduction in the kinetic energy results in the development of a
galactic fountain. When the
resolution is too low to resolve the Jeans scale, the effects of
hydrodynamic drag on the galactic winds will be underestimated.

\section*{Acknowledgements}

We are very grateful to Volker Springel for allowing us to use
\textsc{gadget} and his initial conditions code for the simulations
presented here, as well as for useful discussions and help with his
codes. We gratefully acknowledge discussions with the other members
of the OWLS and Virgo collaborations and we thank the anonymous
referee for a helpful report. The simulations presented here were run
on the Cosmology Machine at the Institute for Computational Cosmology
in Durham as part of the Virgo Consortium research programme and on
Stella, the LOFAR BlueGene/L system in Groningen. This work was
supported by Marie Curie Excellence Grant MEXT-CT-2004-014112.

\end{document}